%% file: entanglement_generation.tex
\documentclass[twocolumn,epsfig,graphics,showpacs,floatfix]{revtex4}

\newcommand{\detector}{*{\xy*+L+<0em,0em>{}="e";"e"+<0em,.4em> **\dir{-};"e"-<0em,.4em> **\dir{-};"e"-<0em,.65em>;"e"+<0em,.2em> *\ellipse :a(0)_{}\endxy} \qw}

\input{declarations}

\usepackage{tikz}
\usepackage{pgfplots}
\usetikzlibrary{shapes, arrows}

\begin{document}

\title{Creation of Entangled Photonic States Using Linear Optics}

\author{Sara Bartolucci}
\author{Patrick M. Birchall}
\author{Mercedes Gimeno-Segovia}
\author{Eric Johnston}
\author{Konrad Kieling}
\email{konrad@psiquantum.com}
\author{Mihir Pant}
\author{Terry Rudolph}
\author{Jake Smith}
\author{Chris Sparrow}
\author{Mihai D. Vidrighin}

\affiliation{PsiQuantum, Palo Alto, USA}

\date\today

\begin{abstract}
  \input{sections/abstract}

\end{abstract}

\maketitle

\section{Entanglement -- the final frontier}
\input{sections/introduction}

\section{Preliminaries}
\input{sections/preliminaries}

\section{The standard Bell state generator}
\label{sec:bsg}
\input{sections/bsg}

\section{Extensions to the BSG}
\label{sec:bsg_extensions}
\input{sections/bsg_extensions}

\section{Bleeding}
\label{sec:bleeding}
\input{sections/bleeding}

\section{Primates}
\label{sec:primates}
\input{sections/primates}

\section{Boosting Fusion}
\label{sec:boosting}
\input{sections/boosting}

\section{Conclusion}
\input{sections/conclusion}

\section{Acknowledgements}
\input{sections/acknowledgements}

\appendix
\input{sections/appendix}

\bibliography{\jobname}{}
\end{document}

%% file: declarations.tex
\usepackage{amsmath,amsfonts,amssymb,graphics,graphicx,epsfig,color,times,bbm}
\usepackage{amsthm}
\usepackage{psfrag}
\usepackage{braket}
\AtBeginDocument{\usepackage{booktabs}}
\graphicspath{{figures/}}
\usepackage{qcircuit}
\usepackage[hidelinks]{hyperref}
\hypersetup{colorlinks=false}

\usepackage{booktabs}
\AtBeginDocument{
  \heavyrulewidth=.08em
  \lightrulewidth=.05em
  \cmidrulewidth=.03em
  \belowrulesep=.65ex
  \belowbottomsep=0pt
  \aboverulesep=.4ex
  \abovetopsep=0pt
  \cmidrulesep=\doublerulesep
  \cmidrulekern=.5em
  \defaultaddspace=.5em
}

\newcommand{\id}{\mathbbm{1}}

\newcommand{\ketbra}[2]{\ket{#1}\!\!\bra{#2}}
\newcommand{\fockmarkup}{}

\makeatletter
\newsavebox{\@brx}
\newcommand{\llangle}[1][]{\savebox{\@brx}{\(\m@th{#1\langle}\)}%
  \mathopen{\copy\@brx\kern-0.5\wd\@brx\usebox{\@brx}}}
\newcommand{\rrangle}[1][]{\savebox{\@brx}{\(\m@th{#1\rangle}\)}%
  \mathclose{\copy\@brx\kern-0.5\wd\@brx\usebox{\@brx}}}
\makeatother

\newcommand{\fket}[1]{\hspace{.1em}|\fockmarkup{#1}\rrangle}
\newcommand{\fbra}[1]{\llangle\fockmarkup{#1}|}
\newcommand{\fketbra}[2]{\fket{#1}\fbra{#1}}

\usepackage[framemethod=tikz]{mdframed}

\definecolor{warning_bgcol}{RGB}{252,248,229}
\definecolor{warning_textcol}{RGB}{111,89,54}
\definecolor{warning_linecol}{RGB}{252,248,229}
\definecolor{danger_bgcol}{RGB}{239,223,222}
\definecolor{danger_textcol}{RGB}{128,60,57}
\definecolor{danger_linecol}{RGB}{239,223,222}
\definecolor{success_bgcol}{RGB}{224,237,216}
\definecolor{success_textcol}{RGB}{68,104,60}
\definecolor{success_linecol}{RGB}{224,237,216}
\definecolor{info_bgcol}{RGB}{220,237,246}
\definecolor{info_textcol}{RGB}{58,100,126}
\definecolor{info_linecol}{RGB}{220,237,246}

\mdfdefinestyle{box_style}{
  skipabove=.7\baselineskip,
  skipbelow=.7\baselineskip,
  innertopmargin=.65\baselineskip,
  innerbottommargin=.65\baselineskip,
  innerleftmargin=.5\baselineskip,
  innerrightmargin=.5\baselineskip,
  splittopskip=1.5\baselineskip,
  splitbottomskip=\baselineskip,
  roundcorner=.3\baselineskip
}
\mdfdefinestyle{warning_style}{
  style=box_style,
  backgroundcolor=warning_bgcol,
  linecolor=warning_linecol,
  fontcolor=warning_textcol,
}
\mdfdefinestyle{success_style}{
  style=box_style,
  backgroundcolor=success_bgcol,
  linecolor=success_linecol,
  fontcolor=success_textcol,
}
\mdfdefinestyle{danger_style}{
  style=box_style,
  backgroundcolor=danger_bgcol,
  linecolor=danger_linecol,
  fontcolor=danger_textcol,
}
\mdfdefinestyle{info_style}{
  style=box_style,
  backgroundcolor=info_bgcol,
  linecolor=info_linecol,
  fontcolor=info_textcol,
}

\newmdenv[style=warning_style]{warning}
\newmdenv[style=info_style]{info}
\newmdenv[style=danger_style]{danger}
\newmdenv[style=success_style]{success}

\newcommand{\Hvec}{{\bf H}}

\newcommand{\modenumber}[1]{\ensuremath{{\scriptstyle #1}\hspace{-1em}}}

%% file: sections/abstract.tex
Using only linear optical elements, the creation of dual-rail photonic entangled states is inherently probabilistic. 
Known entanglement generation schemes have low success probabilities, requiring large-scale multiplexing
to achieve near-deterministic operation of quantum information processing protocols.
In this paper, we introduce multiple techniques and methods to generate photonic entangled states with high probability, which have the potential to reduce the footprint of Linear Optical Quantum Computing (LOQC) architectures drastically.
Most notably, we are showing how to improve Bell state preparation from four single photons to up to $p_{\rm s}=2/3$, boost Type-I fusion to $75\%$ with a dual-rail Bell state ancilla and improve Type-II fusion beyond the limits of Bell state discrimination.

%% file: sections/introduction.tex
The possibility of performing universal quantum computation efficiently using only single photons, linear optics and photon number detection~\cite{KLM01} relies on the interplay between two primitives (i) the generation and manipulation of entanglement by interferometers and (ii) feedforward/adaptivity -- varying the interferometer on unmeasured modes depending on intermediate measurement outcomes.

The paradigm of \emph{Fusion-Based Quantum Computing (FBQC)}~\cite{PsiQ_FBQC} provides a compelling framework for the use of those linear optics primitives for fault-tolerant quantum computation. In that framework, the fundamental primitives needed are:
\begin{itemize}
\item entangling measurements, called \emph{fusions}, which directly correspond to some entangling operations available in linear optics,
\item small, constant-sized entangled states, called \emph{resource states}, which can be built with linear optics by applying sequences of fusions to
   \emph{seed states}, such as Bell states.
\end{itemize}
In the context of linear optics, those seed states in turn are produced by entangling measurements, such as a Bell state generator, acting on single photons. 

The majority of the entangling measurements of interest are probabilistic using linear optics, and while feedforward has not been proven strictly necessary, there are advantages to overcoming the probabilistic nature of the entangling gates in linear optics.
The core issue adaptivity overcomes is that unitary evolution of single photons through interferometers generates only a limited class of entangled states. Measurement on some of the entangled systems (modes) of such states can, however, probabilistically collapse other modes into a much wider class of states. Adaptivity allows for post-selecting events where such measurement is successful. Particularly useful is \emph{multiplexing}, wherein a probabilistic operation is repeated multiple times in parallel and successful events selected out by suitable switching.
In this paper we will present a variety of scenarios where we have found methods for the creation and/or manipulation and/or measurement of useful photonic entanglement that are significantly more effective than previously known protocols. The majority of these rely on judicious use of adaptivity. By ``useful'' entanglement we refer to dual-rail encoded photonic qubit states, and our focus is on stabilizer states and measurements (such as Bell or GHZ states) since these underpin the best understood routes to fault tolerant quantum computing~\cite{PsiQ_FBQC}. Nevertheless, one theme of our results is that making use of states and operations more natural to photons at intermediate steps can be very powerful.
We first consider generation of dual-rail Bell states from four single photons, starting with the standard method in Sec.~\ref{sec:bsg}.
After this, we discuss some extensions to ballistic Bell state generation in Section~\ref{sec:bsg_extensions}, showing that: 

\begin{itemize}
  \item Adding more single photons can slightly improve the success probability.
  \item Using larger interferometers Bell states can be generated using four single photons that are input into random modes. This is particularly useful for Bell state generation from probabilistic sources of (heralded) photons.
\end{itemize}

Despite considerable numerical investigation, the most advanced recent work being that of~\cite{SLMT17} and~\cite{FSK21}, no method for creating Bell states with probability higher than $1/4$ from four single photons and using arbitrary adaptivity was found. In Sec.~\ref{sec:bleeding} we introduce a technique we term ``bleeding'' that uses adaptivity during Bell state generation. Using this we show:
\begin{itemize}
  \item Bell states can be created with probability approaching $2/3$ from four single photons.
\end{itemize}

Linear optical photonic architectures have generally been considered for protocols that involve probabilistic generation of stabilizer states, followed by the use of stabilizer measurements to create even larger states. In Sec.~\ref{sec:primates} we introduce a technique we call ``primates'' that demonstrates the power of using intermediate adaptivity, that is adaptivity involving non-stabilizer states that are specific to photonics. As an example of the potential for this protocol we show: 
\begin{itemize}
  \item Multiplexed GHZ state generation with a reduced number of photon sources compared to known multiplexed qubit-based scheme.
\end{itemize}
   
It is possible to do an entangling CP map on two photonic qubits which destroys one of the qubits -- the entangling Kraus operator takes a generic form $\ketbra{0}{00}+\ketbra{1}{11}$. This operation (termed Type-I fusion in~\cite{BR04}) allows for the creation of larger entangled states from smaller ones. To date it has only been known how to perform Type-I fusion using linear optics in a way such that the probability of success for fusing (i.e., performing a maximally entangling operation on) two initially independent qubit states is $1/2$. In Sec.~\ref{sec:boosted_type1} we show:
\begin{itemize}
  \item The success probability of Type-I fusion on two independent qubits can be boosted to $3/4$ by using an ancillary Bell pair. From this we can deduce that it is possible to create a percolated cluster state ballistically using only Bell states, answering an open question in~\cite{GSBR15}.
\end{itemize}

It is known that there is no method using linear optics to perform a complete Bell basis measurement. It is possible, however, to do a measurement that probabilistically projects onto two elements of a Bell basis or onto product states. Such partial Bell basis measurements can be boosted using two single photon ancillae to a probability of $5/8$, or to success probability $3/4$ with either an ancillary Bell pair~\cite{Grice11} or four ancillary single photons~\cite{EL14}.

Partial Bell basis measurements can also be a useful operation for creating larger entangled states from smaller ones (termed Type-II fusion in~\cite{BR04}), as well as for other protocols such as teleportation. For these uses, however, there is no a-priori need for the measurement to comprise projections onto orthogonal Bell basis states. In particular any POVM with outcomes that are proportional to (possibly non-orthogonal) maximally entangled states is equally useful. In Sec.~\ref{sec:fusion_vs_bsm} we use this insight to show:
\begin{itemize}
  \item Type-II fusion of independent qubits can be achieved with a success probability of $7/12$ using a single photon as a boosting ancilla, and to probability $2/3$ using two single photon ancillae.
\end{itemize}

Each of the topics we analyze can be understood in much more depth and generality than we will present here -- in what follows we discuss only the simplest nontrivial variations that illustrate the key ideas.

%% file: sections/preliminaries.tex
The linear optical schemes we will present below will make use of different ways to
represent logical qubits by states of single photons: \emph{single-rail} means that
the presence (or absence) of a photon in a single optical mode encodes the qubit state
and in \emph{dual-rail} the qubit is represented by the photon being in either of two modes.
To distinguish between, qubit levels and photon numbers, their states are written differently,
for example $\ket{10}=\fket{0110}$ for a pair of dual-rail qubits using the mode pairings $(1,2)$ and $(3,4)$.

Dual-rail encoding acts as an error-detecting code, both loss and multi-photon contamination of a qubit
can be detected by measuring the total number of photons across the two qubit modes locally.
Many examples in the literature
encode dual-rail qubits into the polarization degree of freedom of a single photon, which lends itself naturally to this application.
But it is also well known that any linear optical scheme using polarization can be translated into other degrees of freedom
by appropriate replacements of optical elements. For simplicity of display we will use here an exclusively
path-encoded picture -- all ``modes'' are distinguished by the position of e.g.\ a waveguide on a chip.

All linear optical schemes for generating dual-rail entangled qubits from single photons
rely on obtaining a set of detection patterns upon measurement of a subset of modes; only certain patterns yield the desired state thus making the protocol inherently probabilistic. The success probability of the protocol can be increased by using some of the techniques described in this document, however the inherent stochasticity remains. To bring the entanglement generation scheme closer to an on-demand protocol, we can use \textit{multiplexing}~\cite{kwiatMUX,RMUX} (dubbed muxing). This amounts to repeating a probabilistic protocol multiple times (in space and/or time or exploiting other degrees of freedom such as the photons' frequency), using classical logic to select a successful instance and a reconfigurable switch network to route the outcome to a pre-defined set of modes. There are a number of multiplexing schemes that enable efficient usage of the successful entanglement generation events~\cite{PsiQ_SN}. For the purpose of this paper it is enough to assume that the probability of any entanglement generation scheme described can be pushed to near unity by using an appropriate multiplexing scheme.

A note of caution: we are going to show diagrams describing linear optical circuits
with interferometers depicted not unlike controlled phase gates in the world of qubits.
Besides allowing for very compact diagrams this also allows us to easily follow optical modes through the circuit.

In this convention, horizontal lines are optical modes and beam splitters between two modes are

\centerline{\Qcircuit @C=1em @R=1em { & \control \qw & \rstick{} \qw \\\lstick{} & \ctrl{-1} & \rstick{} \qw}\vspace{1em}}
\noindent where ``reflection'' occurs when photons move from the lower mode to the upper one (or vice versa) and ``transmission``
occurs when photons stay in the same mode.
Those are 50:50 beam splitters and if not mentioned otherwise, we will assume
real, symmetric ones (referring to their real and symmetric transfer matrices).
They are described by a $2\times 2$ Hadamard matrix $H^{(2)}$ -- the superscript indicating the dimensionality.
We will use the convention that the modes as shown in the diagrams index the rows and columns of the
corresponding transfer matrix ($1$-indexed) from top to bottom.

Similarly, interferometers with unbiased transition amplitudes between more than two modes may be shown as 

\centerline{\Qcircuit @C=1em @R=1em {
  & \control \qw & \qw \\
  & \ctrl{-1} & \qw \\
  & \ctrl{-2} & \qw \\
  & \ctrl{-3} & \qw
}\vspace{1em}}
\noindent where the default will be a $4\times 4$ Hadamard-type transfer matrix $H^{(4)}$.
Depending on the context this may be describing a
discrete Fourier transform (DFT) transfer matrix as well (but we'll make sure to mention when the difference is really
relevant).

The presentation below will be mainly focussing on the ``ideal'' behavior -- in the absence of imperfections and losses.
In this context, the operation of a state generation or fusion circuit is defined by input states, the transfer matrix of
linear optical elements and photon numbers in the detectors (``detection patterns''). In contrast, the actual physical
setup used to implement this transfer matrix is of no relevance. Note that this is a simplification, as
errors in the outcomes will depend strongly on e.g.\ the physical realizations of beam splitters or the specific
decomposition used to implement an arbitrary transfer matrix with a fixed set of primitive phases and beam splitters.

%% file: sections/bsg.tex
\begin{figure}
  \input{figures/bsg_standard.tex}
  \caption{
    Schematics of a Bell state generator. $m$ are the measurements acting on the ancilla modes (thus entangling four pairs of $(\fket{01}+\fket{10})/\sqrt{2}$)
    and $M$ the measurements on the signal modes (initialized in the state $\fket{1}^{\otimes 4}$).
    \label{fig:bsg_standard}
  }
\end{figure}
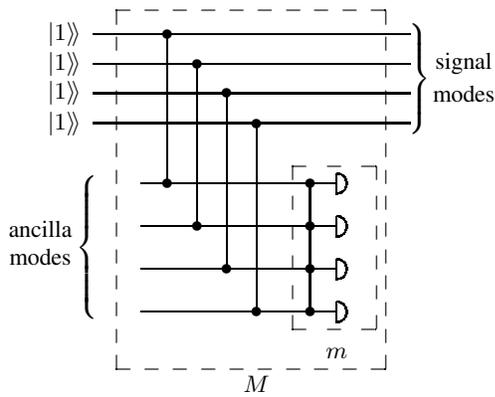

A universal example for discussing some of the effects in the following sections is a Bell state generator (BSG).
Among the many BSG variations, we will consider one amenable to 
modifications that will be introduced later (see Fig.~\ref{fig:bsg_standard}).
The operation of this BSG is simple -- four unentangled single photons are injected into the device and, depending on probabilistic detection events involving two photons, a dual-rail Bell state is generated at the output. 
It is based on~\cite{Zhang08}, freed of any polarization. We are presenting its operation in detail,
which will allow us to discuss extensions and improvements easily later on.

\subsection{Measurement Operators}
\label{sec:bsg_operators}
\begin{figure}
  \input{figures/single_mode.tex}
  \caption{
    Photon subtraction primitive used to  subtract a photon from a signal mode.
    The ancilla mode is in a vacuum state before interacting at a 50:50 beam splitter with the signal.
    The boxes labeled $M$ and $m$ enclose the measurement devices described by
    Eq.~(\ref{eqn:single_mode}).
    We will refer to the beam splitter between signal and ancilla as {\it coupling beam splitter}.
    \label{fig:single_mode}
  }
\end{figure}
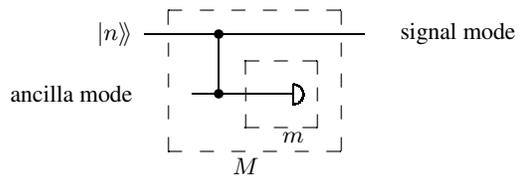

In order to illustrate the principles at work in the BSG, we will first consider the toy model shown in Fig.~\ref{fig:single_mode}.

Here, the collection $\bra{m_{(i)}}$ describe the projective measurement due to the photon number resolving detector
on the ancilla mode and $\hat M_{(i)}$ the {\it measurement operators} (Kraus operators)
representing the effect of the entire device acting on the signal mode.
Here we are using the convention that indices in brackets refer to detection patterns.
The detections of zero and single photons are described by
\begin{equation}\begin{array}{rclcrcl}
  \hat M_{(0)} &=& 2^{-\hat n/2}  &\quad& \bra{m_{(0)}} &=& \fbra{\rm 0} \\
  \hat M_{(1)} &=& \hat A  \, 2^{-\hat n/2}  &\quad& \bra{m_{(1)}} &=& \fbra{1} = \fbra{\rm 0} \hat a
\end{array}\label{eqn:single_mode}\end{equation}
where $\hat n$ is the signal mode's photon number operator and $\hat A$ and $\hat a$ the annihilation operators on the signal and ancilla modes, respectively.

The BSG consists of four copies of the photon subtraction primitive introduced above, 
connected via a 4-mode Hadamard interferometer. $\bra{m}$ can be generalized as per the outline in 
Fig.~\ref{fig:bsg_standard} by {\it backpropagating} the
annihilation operators used to describe the photon detectors through the four-mode Hadamard.
Some examples of these states projected onto are
\begin{eqnarray*}
  \bra{m_{(1100)}}\! &=&\! \fbra{0000} (\Hvec_1\cdot {\bf \hat a})(\Hvec_2\cdot {\bf \hat a}) \\
          &=&\!2^{-1/2}\bra{\phi^-} + 2^{-3/2}\bigl(\fbra{2000} \\
           &&\!-\, \fbra{0200}+\fbra{0020}-\fbra{0002}\bigr) \\
  \bra{m_{(2000)}} &=& \fbra{0000} (\Hvec_1\cdot {\bf \hat a})^2 / \sqrt 2 \\
          &=& \sqrt{3}/2 \bra{W_{4,2}} + 2^{-2}\bigl(\fbra{2000} \\
           && +\, \fbra{0200}+\fbra{0020}+\fbra{0002} \bigr)
\end{eqnarray*}
where ${\bf \hat a}$ is the vector of annihilation operators on the ancilla modes, $\Hvec_i$ 
denotes the $i$-th row vector of a real, normalized Hadamard matrix $H^{(4)}$, and
$\ket{W_{4,2}}$ is a $W$-type state of $2$ photons on $4$ modes as defined in Appendix~\ref{sec:h4_projection}.
$\ket{\phi^-}$ is the Bell state on dual-rail qubits defined on modes $(1,2)$ and $(3,4)$.
We can now picture the
BSG as a projective measurement acting on a collection of {\it single-rail Bell states}
$\ket{\psi_1}^{\otimes 4}$ (with $\ket{\psi_1}=(\fket{10}+\fket{01})/\sqrt{2}$)
created by the four initial single photons $\fket{1}^{\otimes 4}$ and the coupling beam splitters.

The corresponding operators $\hat M$ acting on the $\fket{1111}$ input in the signal modes
are
\begin{eqnarray}
  \hat M_{(1100)} &=& (\Hvec_1\cdot {\bf \hat A})(\Hvec_2\cdot {\bf \hat A}) \, 2^{-\hat n/2} \label{eqn:bsg_subtraction} \\
  \hat M_{(2000)} &=& (\Hvec_1\cdot {\bf \hat A})^2 \, 2^{-(\hat n + 1)/2} \nonumber
\end{eqnarray}
where ${\bf \hat A}$ is the vector of annihilation operators on the signal modes and $\hat n$ is the number operator for the total number of signal photons.
Note that these measurement operators are still described by products corresponding to two independent photon subtractions from the signal modes.

These operators can be simplified if we constrain each signal mode to contain at most a single photon. We truncate 
$\hat M$ and $\bra{m}$ by applying the projection to the $\{\fket{0},\fket{1}\}$-subspace, $\hat P=(\fketbra{0}{0} + \fketbra{1}{1})^{\otimes 4}$:
\begin{eqnarray*}
  \bra{m'_{(1100)}} &= \bra{m_{(1100)}}\hat P &= \frac{1}{\sqrt 2}\bra{\phi^-} \\
  \hat M'_{(1100)}  &= \hat M_{(1100)}\hat P  &= \frac{1}{8}(\hat A_1\hat A_3-\hat A_2\hat A_4) \\
  \bra{m'_{(2000)}} &= \bra{m_{(2000)}}\hat P &= \frac{\sqrt 3}{2} \bra{W_{4,2}} \\
  \hat M'_{(2000)}  &= \hat M_{(2000)}\hat P  &= \frac{1}{8\sqrt{2}}(\hat A_3\hat A_4 + \hat A_2\hat A_4 + \hat A_2\hat A_3 \\
    &&\ + \hat A_1\hat A_4 + \hat A_1\hat A_3 + \hat A_1\hat A_2) \\
\end{eqnarray*}

By applying the $\hat M'$ to the single-photon input states, we get for example
\begin{eqnarray*}
  \hat M'_{(1100)}\fket{1111} &=& -2^{-5/2}\ket{\phi^-} \\
  \hat M'_{(2000)}\fket{1111} &=& \sqrt{3}/8\ket{W_{4,2}}
\end{eqnarray*}

\subsection{Success probabilities}
\label{sec:bsg_probs}
The probability of two photons ending up in the first two detectors is
\[
  p(1100) = \lVert \hat M'_{(1100)}\fket{1111}\rVert^2 = 1 / 32
\]
All of the six $\hat M'$ corresponding to two single photon detections cause production of maximally entangled states with the same probability, resulting in an
overall probability of $3/16$. The four events where two photons are detected in the same ancilla mode can also occur with the same overall probability, 
\[
  p(2000) = \lVert \hat M'_{(2000)}\fket{1111}\rVert^2 = 3 / 64
\]
in which case the measurement operators $\hat M'$ produce non-maximally entangled $W$-type states.
An intuitive explanation goes as follows:

In order to herald a two-qubit state in the signal modes, two photons need to be detected.
After the coupling beam splitters, there are $2^4$ equally probable terms,
of which ${4\choose 2}=6$ have two photons in the detector modes.
They give us an upper bound on the probability of successfully
generating a Bell state of $6/2^4=3/8$.
Given an equal superposition of these six terms after the coupling beam splitters,
the maximally entangled $\bra{m'}$ amount to 50\% of the detection events -- the $3/16$ we saw before.

\subsection{Corrections}
\label{sec:bsg_corrections}
While there are six patterns indicating generation of maximally entangled qubit states, not all of them are equal:
Being totally symmetric with respect to the four signal modes, the BSG does not distinguish any mode pairings for dual-rail
encoding and can be expected to produce a dual-rail Bell state in all six possible mode permutations.
For any given choice of assigning the four output modes to two dual-rail qubits, those six permutations are manifested in the fact that the BSG
produces
\begin{eqnarray*}
  \pm\ket{\phi^-} &=& \pm(\fket{0101}-\fket{1010})/\sqrt{2} \\
  \pm\ket{\psi^-} &=& \pm(\fket{0110}-\fket{1001})/\sqrt{2}
\end{eqnarray*}
(the qubit Bell states) and
\begin{eqnarray*}
  \pm\ket{\chi^-} &=& \pm(\fket{1100}-\fket{0011})/\sqrt{2}
\end{eqnarray*}
(a non-qubit maximally entangled state of two photons on four modes).
Depending on operations downstream
those outcomes may need to be cast into one specific choice of the three types of states above by applying
mode swaps (equivalent to a bit flip if the qubit mode allocation does not change).

This feed-forward correction depends on the detection outcomes of the BSG but may be subsumed in later operations
without the need for additional active switching elements, as is the case with Pauli corrections in many applications.
There are examples in~\cite{PsiQ_SN} of how downstream MUX networks can be exploited for this purpose.

\subsection{Distillation}
As discussed above, half of the two-photon detection events (where two photons end up in a single detector)
produce -- up to local phases -- non-maximally entangled $W$-type states (see Appendix~\ref{sec:h4_projection}).

Seen as (symmetrized) two-partite states, it is clear that a procrustean distillation scheme
can be used to yield maximally entangled states~\cite{joo2007one,Kieling08}:
As any quadratic form can be turned into a sum of squares by a linear (even unitary) transformation,
a linear optical network can be found to turn $\ket{W_{4,2}}$ into the Schmidt decomposition
\begin{eqnarray*}
  && \sqrt{3/4}\fket{2000} + \sqrt{1/12}\fket{0200} \\
  && + \sqrt{1/12}\fket{0020} + \sqrt{1/12}\fket{0002}
\end{eqnarray*}
All coefficients can be made equal by damping the first Schmidt coefficient by coupling a vacuum ancilla
to the first mode and then post-selecting on detecting vacuum in it.
Applying another linear optical network allows to turn those states back into dual-rail qubits,
more specifically maximally entangled Bell pairs.

The probability of this process is dictated by the amplitude lost during damping -- $1/3$ in this case.
Note that while the distillation network will depend on the local phases of the $W$-type states,
all of those states may be distilled to Bell states with the same probability.
Thus, distillation can be used to improve the overall success probability of the BSG to
\begin{equation}
  \frac{3}{8}\left(\frac{1}{2} + \frac{1}{2} \times \frac{1}{3}\right) = \frac{1}{4}. \nonumber
\end{equation}

\subsection{Avenues for generalization}
The formalism depicted above allows for a variety of interpretations, each amenable to different generalisations and enhancements in the sections to follow:
\begin{itemize}
  \item Steering on maximally entangled states $\ket{\psi_1}$ using projective measurements described by the $\bra{m}$ as shown in Eqs.~(\ref{eqn:povm1}-\ref{eqn:povm2})).
  \item Two-fold photon subtraction of photons spread evenly over four modes from $\fket{1}^{\otimes 4}$ as described by $M$ in Eq.~(\ref{eqn:bsg_subtraction}).
     This interpretation shows the way towards ``bleeding'' (Sec.~\ref{sec:bleeding}).
  \item Four-partite ``fusion'' of single-rail Bell states $\ket{\psi_1}$ (or just think of PEPS) with the projective measurements $\bra m$.
     Bell states can also be represented as outcomes of a four-way fusion on different states, with an enhanced success probability (Sec.~\ref{sec:bsg_8photons}).
     The same single-rail four-way fusion applied to one rail each of four dual-rail Bell pairs results in 6-GHZ states with the same probability of $p=3/16$.
  \item A similar point of view is to interpret Bell state generation as a kind of ``teleported'' measurement of $\bra{m'}$. As is the case with all
     teleportations, the resource entanglement (here embodied by the four single-rail bell pairs) is consumed.
     Enhancing the probability of performing $\bra{m'}$ leads to improvements in Bell state generation (Sec.~\ref{sec:bsg_boosting}).
\end{itemize}

%% file: figures/bsg_standard.tex
\centerline{\Qcircuit @C=1em @R=1em {
  \lstick{\fket{1}} & \qw & \qw & \control \qw & \qw & \qw & \qw & \qw & \qw & \qw & \qw & \qw & \qw \\
  \lstick{\fket{1}} & \qw & \qw & \qw & \control \qw & \qw & \qw & \qw & \qw & \qw & \qw & \qw & \qw & \rule{2.5em}{0em}\mbox{signal} \\
  \lstick{\fket{1}} & \qw & \qw & \qw & \qw & \control \qw & \qw & \qw & \qw & \qw & \qw & \qw & \qw & \rule{2.5em}{0em}\mbox{modes} \\
  \lstick{\fket{1}} & \qw & \qw & \qw & \qw & \qw & \control \qw & \qw & \qw & \qw & \qw & \qw & \qw {\gategroup{1}{13}{4}{13}{.75em}{\}}} \\
  & & & & & & & & & & & \\
  & & & \ctrl{-5} & \qw & \qw & \qw & \qw & \control \qw &\detector & \\
  & & & \qw & \ctrl{-5} & \qw & \qw & \qw & \control \qw &\detector & \\
  & & & \qw & \qw & \ctrl{-5} & \qw & \qw & \control \qw &\detector & \push{\rule{.1em}{0em}} \\
  & & & \qw & \qw & \qw & \ctrl{-5} & \qw & \ctrl{-3} \qw & \detector\gategroup{6}{9}{9}{11}{1.2em}{--}& \\
  & & & & & & & & & \rule{0em}{1.1em} m  \gategroup{1}{3}{10}{11}{2em}{--} & & & & & \\
  & & & & & & \rule{0em}{2em} M
  \inputgroupv{6}{9}{.5em}{2.5em}{\parbox{5em}{ancilla \\ modes}\quad}
}\vspace{2em}}

%% file: figures/single_mode.tex
\centerline{\Qcircuit @C=1em @R=1em {
  \lstick{\fket{n}} & \qw & \qw & \control \qw & \qw & \qw & \qw & \qw & \qw & \qw & \rstick{\mbox{signal mode}}\\
  & & & & & & & & & & & \\
  \lstick{\mbox{ancilla mode}} & & & \ctrl{-2} & \qw & \qw &\detector &\\
  & & & & & & \rule{0em}{1.3em}m  \gategroup{3}{6}{3}{7}{2em}{--} & & & & & \\
  & & & & \rule{0em}{2em}M  \gategroup{1}{3}{4}{8}{2em}{--} & & & & & \\
}\vspace{2em}}

%% file: sections/bsg_extensions.tex
\subsection{Boosting}
\label{sec:bsg_boosting}
As we can see in Eqs.~(\ref{eqn:povm1}-\ref{eqn:povm2}), four of the six ``success'' outcomes correspond to projections onto
dual-rail Bell states. In fact, $m$ in Fig.~\ref{fig:bsg_standard} {\it is} a probabilistic Bell state measurement on the subspace
where the input modes are occupied by dual-rail qubit states.
``Boosting'' of the success probability of those types of state discrimination has been
previously shown in~\cite{Grice11,EL14,GSBR15}. In contrast to those state discrimination tasks,
in the BSG context, the circuit acts on a larger space,
effectively projecting onto the dual-rail two-qubit space {\it and} performing a BSM on those qubits.

Replacing $m$ by e.g.\ the single-photon boosted Bell state discriminator from~\cite{EL14}
(which improves Bell state discrimination from $50\%$ to $75\%$), the success probability
increases to $7/32=21.875\%$ (with four ancilla photons) or $13/64=20.3125\%$ (with two ancilla photons).

This comes at the cost of adding another state $\ket{\chi^+}=2^{-1/2}(\fket{1100}+\fket{0011})$ to the set of possible output states.
While the improvement looks like a small one compared to selecting a Bell state from two parallel BSGs,
it shows a simple example of how results from state discrimination can be exploited in the context of state preparation.

\subsection{More photons}
\label{sec:bsg_8photons}
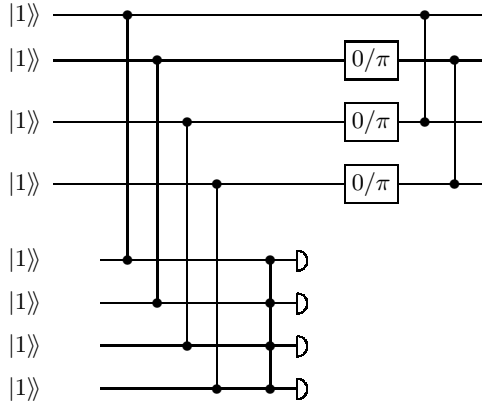
\begin{figure}
  \input{figures/bsg_8photon.tex}
  \caption{
    Bell state generator using $8$ single photon inputs. All detection patterns of $6$ photons
    indicate successful creation of a maximally entangled state of $2$ photons in the $4$ signal modes.
    The $0$/$\pi$ phase shifters are set dependent on outcomes of the detectors in the ancilla modes.
    \label{fig:bsg_8photon}
  }
\end{figure}

Bell states can also be generated using four $\fket{20}+\fket{02}$ resources (instead of single-rail Bell states),
obtained by injecting single photons in {\em all} eight input modes of the BSG circuit (see Fig.~\ref{fig:bsg_8photon}).
Similar to how subtracting three photons from a state $\bigl(\fket{10}+\fket{01}\bigr)^{\otimes 4}$ leaves a single photon in a
balanced superposition of being in one of four modes,
detecting six photons here results in a state
$\fket{2000}+\fket{0200}+\fket{0020}+\fket{0002}$ (up to phases depending on the exact detection pattern).
These states can be converted into Bell states by applying (1) corrective phase shifts that depend on detection outcomes
and (2) two beam splitters between pairs of the signal modes.
Such a $6$-photon detection results from one bunched two-photon term staying in the signal modes while the other three are
detected in the ancilla modes. Thus, the probability of getting maximally entangled states is the same as choosing
one bunched pair out of four -- 25\%.

\subsection{Random inputs}
A description of the BSG by $\bra{m^\prime}$ acting on $\ket{\psi_1}^{\otimes 4}$ is independent of how
these input states are produced.
So far we have considered those $\ket{\psi_1}$ as outcomes from single photons injected into
the signal modes of the coupling beam splitters. In fact, sending a photon into an ancilla mode of a coupling
beam splitter also generates a $\ket{\psi_1}$, albeit up to a flip of the relative phase.
The changed relative phase of any of the inputs will translate into a potential flip of the generated Bell states' phases.

Depending on the physical platform used for implementation,
preparation of single photons could be a probabilistic process, e.g.\ multiplexed pumped \& heralded
pair sources. Thus, instead of considering only a single input state, it can be helpful to allow for any of the
$2^4=16$ possible inputs to the BSG where one single photon enters each coupling beamsplitter either in the signal {\it or} the ancilla mode.
Instead of having to multiplex a number of sources down to a single mode, it is sufficient to output the photon in either of two modes,
thus potentially reducing the size of the multiplexing network. This comes only at the cost of having to correct
for phases in the produced Bell state -- something that can be taken care of by already existing active elements
in the qubit path downstream.

\subsection{Accepting even more randomness}
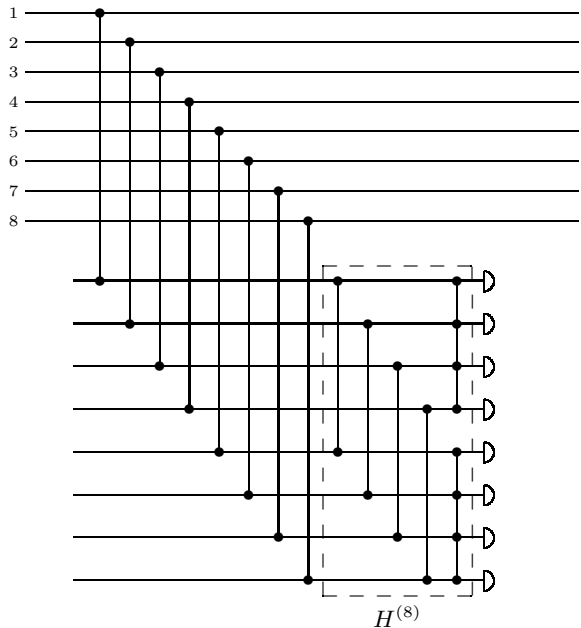
\begin{figure}
  \input{figures/random_input_bsg.tex}
  \caption{
    Duplicating the BSG circuit and adding pairwise beam splitters between the ancilla modes leads to circuit of
    eight coupling beam splitters between signal and ancilla followed by an eight-mode Hadamard on the ancilla.
    \label{fig:random_input_bsg}
  }
\end{figure}

We can further expand the BSG by duplicating the complete network and adding four 50:50 beam splitters between the network's ancilla modes, as shown in Fig.~\ref{fig:random_input_bsg}.

It is worth noticing that the overall interferometer on the ancilla modes is now an 8-mode Hadamard (described by a transfer matrix $H^{(8)}$).

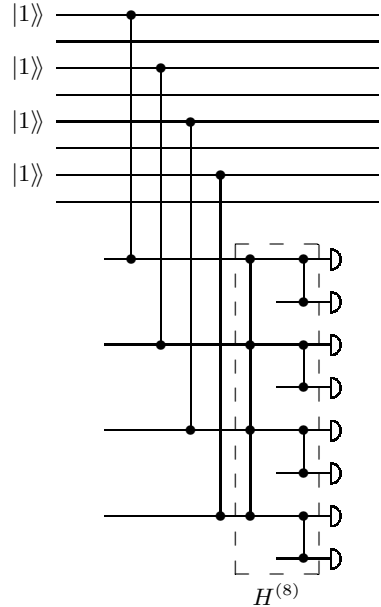
\begin{figure}
  \input{figures/random_input_bsg_decomposed.tex}
  \caption{
    Input of single photons in $4$ of the signal modes may lead to those photons effectively "seeing" a BSG with
    some additional detector fan-out.
    \label{fig:random_input_bsg_decomposed}
  }
\end{figure}

Now there are $8$ beamsplitters coupling the signal to the ancilla and we can inject single photons in any $4$ of them to produce Bell states. Depending on which modes the photons enter,
there are two possible situations:
\begin{itemize}
  \item There exists a decomposition of $H^{(8)}$ into $H^{(4)}$ and additional beam splitters $H^{(2)}$,
    such that it is easy to identify a standard 8-mode BSG circuit in the forward light cone of those photons (see Fig.~\ref{fig:random_input_bsg_decomposed}).
    For those input patterns the same measurement operators as in Sec.~\ref{sec:bsg_operators} are applied and Bell states are produced in the four chosen input modes
    with $p=3/16$. This happens for all patterns where the bit-wise sum (mod 2) of the input mode numbers (minus 1) adds up to $0$, $(1,4,6,7)$ being an example.  
  \item There exists no such decomposition, in which case Bell states are still produced, albeit with a 50\% reduced success probability.
    An example for a configuration of input modes where this happens is $(1,5,6,7)$.
\end{itemize}

Every input pattern of four single photons can thus be used to generate Bell states.
This construction can be generalised to yield BSG circuits with $2^k$ signal modes.

Depending on the process used to generate single photons, a circuit that forgiving may help in reducing the size of the multiplexing networks
required. The trade-off here is that the resulting Bell states will populate a random subset of $4$ of the $>4$ signal modes.
Having said that, for a given set of input modes, the output modes are fixed -- the result is produced
in the same four signal modes occupied by the initial single photons.
If it is required by an application that the Bell state has support on four specific modes, this
can be achieved easily using one of the switch networks presented in~\cite{PsiQ_SN}.

It is important to note that, as it is the case with the basic Bell state generator described in Sec.~\ref{sec:bsg_corrections},
different detection patterns lead to Bell states that may need mode swap and/or phase corrections.

%% file: figures/bsg_8photon.tex
\centerline{\Qcircuit @C=1em @R=1em {
  \lstick{\fket{1}} & \qw & \qw & \control \qw & \qw & \qw & \qw & \qw & \qw & \qw & \qw & \qw & \ctrl{2} & \qw & \qw \\
  \lstick{\fket{1}} & \qw & \qw & \qw & \control \qw & \qw & \qw & \qw & \qw & \qw & \qw & \gate{0 / \pi} & \qw & \ctrl{2} & \qw \\
  \lstick{\fket{1}} & \qw & \qw & \qw & \qw & \control \qw & \qw & \qw & \qw & \qw & \qw & \gate{0 / \pi} & \control \qw & \qw & \qw \\
  \lstick{\fket{1}} & \qw & \qw & \qw & \qw & \qw & \control \qw & \qw & \qw & \qw & \qw & \gate{0 / \pi} & \qw & \control \qw & \qw \\
  & & & & & & & & & & & \\
  \lstick{\fket{1}} & & & \ctrl{-5} & \qw & \qw & \qw & \qw & \control \qw & \detector \\
  \lstick{\fket{1}} & & & \qw & \ctrl{-5} & \qw & \qw & \qw & \control \qw & \detector \\
  \lstick{\fket{1}} & & & \qw & \qw & \ctrl{-5} & \qw & \qw & \control \qw & \detector \\
  \lstick{\fket{1}} & & & \qw & \qw & \qw & \ctrl{-5} & \qw & \ctrl{-3} \qw & \detector
}}

%% file: figures/random_input_bsg.tex
\centerline{\Qcircuit @C=1em @R=1em {
  \modenumber{1} & & \qw & \qw & \control \qw & \qw & \qw & \qw & \qw & \qw & \qw & \qw & \qw & \qw & \qw & \qw & \qw & \qw & \qw & \qw & \qw & \qw \\
  \modenumber{2} & & \qw & \qw & \qw & \control \qw & \qw & \qw & \qw & \qw & \qw & \qw & \qw & \qw & \qw & \qw & \qw & \qw & \qw & \qw & \qw & \qw \\
  \modenumber{3} & & \qw & \qw & \qw & \qw & \control \qw & \qw & \qw & \qw & \qw & \qw & \qw & \qw & \qw & \qw & \qw & \qw & \qw & \qw & \qw & \qw \\
  \modenumber{4} & & \qw & \qw & \qw & \qw & \qw & \control \qw & \qw & \qw & \qw & \qw & \qw & \qw & \qw & \qw & \qw & \qw & \qw & \qw & \qw & \qw \\
  \modenumber{5} & & \qw & \qw & \qw & \qw & \qw & \qw & \control \qw & \qw & \qw & \qw & \qw & \qw & \qw & \qw & \qw & \qw & \qw & \qw & \qw & \qw \\
  \modenumber{6} & & \qw & \qw & \qw & \qw & \qw & \qw & \qw & \control \qw & \qw & \qw & \qw & \qw & \qw & \qw & \qw & \qw & \qw & \qw & \qw & \qw \\
  \modenumber{7} & & \qw & \qw & \qw & \qw & \qw & \qw & \qw & \qw & \control \qw & \qw & \qw & \qw & \qw & \qw & \qw & \qw & \qw & \qw & \qw & \qw \\
  \modenumber{8} & & \qw & \qw & \qw & \qw & \qw & \qw & \qw & \qw & \qw & \control \qw & \qw & \qw & \qw & \qw & \qw & \qw & \qw & \qw & \qw & \qw \\
  & & & & & & & & & & & & \\
  & & & & \ctrl{-9} & \qw & \qw & \qw & \qw & \qw & \qw & \qw & \control \qw & \qw & \qw & \qw & \control \qw &\detector & \\
  & & & & \qw & \ctrl{-9} & \qw & \qw & \qw & \qw & \qw & \qw & \qw & \control \qw & \qw & \qw & \control \qw &\detector & \\
  & & & & \qw & \qw & \ctrl{-9} & \qw & \qw & \qw & \qw & \qw & \qw & \qw & \control \qw & \qw & \control \qw &\detector & \\
  & & & & \qw & \qw & \qw & \ctrl{-9} & \qw & \qw & \qw & \qw & \qw & \qw & \qw & \control \qw & \ctrl{-3} \qw & \detector \\
  & & & & \qw & \qw & \qw & \qw & \ctrl{-9} & \qw & \qw & \qw & \ctrl{-4} & \qw & \qw & \qw & \control \qw &\detector & \\
  & & & & \qw & \qw & \qw & \qw & \qw & \ctrl{-9} & \qw & \qw & \qw & \ctrl{-4} & \qw & \qw & \control \qw &\detector & \\
  & & & & \qw & \qw & \qw & \qw & \qw & \qw & \ctrl{-9} & \qw & \qw & \qw & \ctrl{-4} & \qw & \control \qw &\detector & \\
  & & & & \qw & \qw & \qw & \qw & \qw & \qw & \qw & \ctrl{-9} & \qw & \qw & \qw & \ctrl{-4} & \ctrl{-3} \qw & \detector & \\
  & & & & & & & & & & & & & &  \rule{0em}{1.1em}H^{(8)}  \gategroup{10}{13}{17}{17}{1em}{--} & & & & &
}\vspace{2em}}

%% file: figures/random_input_bsg_decomposed.tex
\centerline{\Qcircuit @C=1em @R=1em {
  \lstick{\fket{1}} & \qw & \qw & \control \qw & \qw & \qw & \qw & \qw & \qw & \qw & \qw & \qw & \qw \\
  & \qw & \qw & \qw & \qw & \qw & \qw & \qw & \qw & \qw & \qw & \qw & \qw \\
  \lstick{\fket{1}} & \qw & \qw & \qw & \control \qw & \qw & \qw & \qw & \qw & \qw & \qw & \qw & \qw \\
  & \qw & \qw & \qw & \qw & \qw & \qw & \qw & \qw & \qw & \qw & \qw & \qw \\
  \lstick{\fket{1}} & \qw & \qw & \qw & \qw & \control \qw & \qw & \qw & \qw & \qw & \qw & \qw & \qw \\
  & \qw & \qw & \qw & \qw & \qw & \qw & \qw & \qw & \qw & \qw & \qw & \qw \\
  \lstick{\fket{1}} & \qw & \qw & \qw & \qw & \qw & \control \qw & \qw & \qw & \qw & \qw & \qw & \qw \\
  & \qw & \qw & \qw & \qw & \qw & \qw & \qw & \qw & \qw & \qw & \qw & \qw \\
  & & & & & & & & & & & \\
  & & & \ctrl{-9} & \qw & \qw & \qw & \control \qw & \qw & \control \qw & \detector & \\
  & & & & & & & & & \ctrl{-1} & \detector & \\
  & & & \qw & \ctrl{-9} & \qw & \qw & \control \qw & \qw & \control \qw & \detector & \\
  & & & & & & & & & \ctrl{-1} & \detector & \\
  & & & \qw & \qw & \ctrl{-9} & \qw & \control \qw & \qw & \control \qw & \detector & \\
  & & & & & & & & & \ctrl{-1} & \detector & \\
  & & & \qw & \qw & \qw & \ctrl{-9} & \ctrl{-6} & \qw & \control \qw & \detector & \\
  & & & & & & & & & \ctrl{-1} & \detector & \\
  & & & & & & & & \rule{0em}{1.1em}H^{(8)}  \gategroup{10}{8}{17}{10}{1em}{--} & & & & &
}\vspace{2em}}

%% file: sections/bleeding.tex
\subsection{Deterministic photon subtraction}
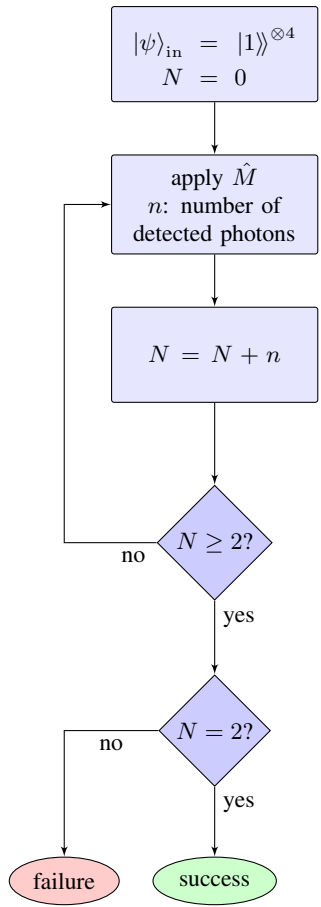
\begin{figure}
  \input{figures/bleeding_protocol.tex}
  \caption{
    Protocol for bleeding off two photons from a four-photon input state:
    Repeated attempts to subtract photons (``stages'') finish when the total number of detected photons $N$ reach (success)
    or exceed (failure)~$2$.
    \label{fig:bleeding_protocol}
  }
\end{figure}

In Sec.~\ref{sec:bsg_probs} we saw that there is an upper limit to the BSG success probability of $3/8$ based on the consideration
that exactly $2$ photons need to end up in the detector modes.
But this is {\it not} where the BSG story ends.
In fact, the probability of measuring in the $2$ photon subspace can be pushed from $3/8$ to $1$
(and conditional measurement probabilities for specific patterns scaling accordingly) without using any additional photons.
All this requires is the ability to perform feed-forward, essentially to repeat ``weak'' photon subtraction until two photons
are detected -- a process we will refer to as ``bleeding'' of two single photons from the signal.

The ability to implement the bleeding protocol follows from two observations:
Firstly, we note that the measurement operators in Eq.~(\ref{eqn:bsg_subtraction}) are products of commuting operators of the form $\Hvec_i\cdot \bf \hat A$.
In fact, $\hat M_{(1000)}$ is the action of $M$ (Fig.~\ref{fig:bsg_standard}) given the detection of a {\it single} photon.
By applying $M$ twice and post-selecting on single photon detection events in each of the two stages,
we end up with the same operator (up to normalization) as if we detected both photons in a single stage. 
Secondly, we note that upon detection of no photons at all, the state in the signal mode is not changed.
It is merely ``damped'' due to the finite success probability of this event occurring: $\hat M_{(0000)}=2^{-\hat n/2}$.
$\hat M_{(0000)}$ can be applied arbitrarily often and commutes with any single-photon subtraction 
operators
\footnote{Those operators are commuting in the sense that the {\it normalized} output state does not depend on the order those operators
are applied in. But normalization depends on it and so does the probability of getting outcomes in a specific order.}, such as $\hat M_{(1000)}$. 
With these ingredients, we can construct the algorithm presented in Fig.~\ref{fig:bleeding_protocol}.

By making the beamsplitter reflectivities that couple the signal to the detector modes smaller,
detections of higher photon numbers are suppressed.
For the input state $\fket{1111}$, single-photon events are suppressed linearly
(probability of $\hat M_{(1000)}$ now scaling with $r(1-r)^3$),
two-photon events are suppressed quadratically (probability of $\hat M_{(1100)}$ scaling with $r^2(1-r)^2$) and so on.
In the limit of infinitesimal coupler reflectivities (and potentially infinitely many stages),
there is virtually no risk of subtracting multiple photons in a single stage.
Zero-photon events on the other hand, which become the dominating outcomes 
(the probability of $\hat M_{(0000)}$ scaling with $(1-r)^4$),
are removed from the consideration due to subsequent retries --
the only remaining operations that change the signal state being the single photon subtraction events.
This retry-until-success mode of operation
\footnote{In contrast to other repeat-until-success protocols in linear optics, here subsequent attempts
act on the same set of photons.}
allows us to re-attempt the ``bleeding'' of single photons until exactly two are subtracted.

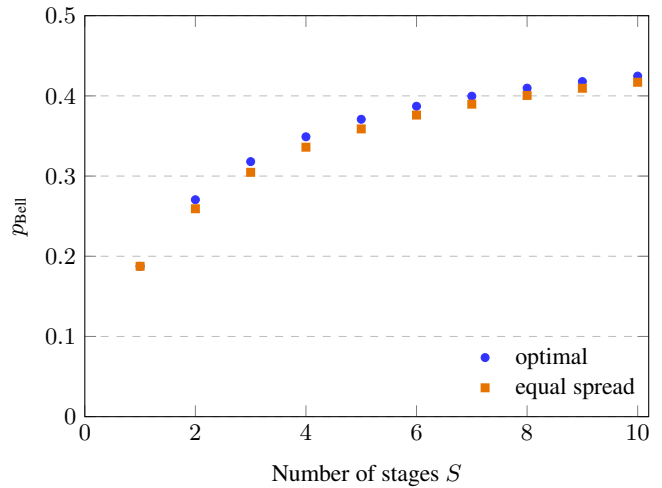
\begin{figure}
  \input{figures/bleeding_stages.tex}
  \caption{
    Change of success probability with increased number of ``bleeding'' stages (without distillation). Coupler reflectivities for an
    equal ``spread'' over all stages are given by Eq.~(\ref{eqn:equal_spread}).
    The point at $S=1$ corresponds to the standard Bell state generator $p(S=1)=3/16$ and both choices
    of reflectivities converge to $p(S\rightarrow\infty)=1/2$.
    \label{fig:bleeding_stages}
  }
\end{figure}

The probability of succesfully bleeding two photons in a given number of stages $S$ depends on the exact choice of reflectivities,
and so does $\lim_{S\rightarrow\infty}p$.
A natural choice is to spread an equal amount of the initial amplitude over all detector stages and the signal,
\begin{equation}
  r_k = \frac{1}{S-k+2}
  \label{eqn:equal_spread}
\end{equation}
with $k=1,\ldots,S$ being the stage index.
Summing up (see Appendix~\ref{sec:bleeding_probs} for details) the probabilities for all possible two-photon events results in (see Fig.~\ref{fig:bleeding_stages})
\[
  p(S) = \frac{S (1 + S + S^2)}{(1 + S)^3}
\]
and
\[
  \lim_{S\rightarrow\infty}p(S) = 1 \,.
\]
As we have seen earlier, within those two-photon patterns half herald a Bell state and the other half $\ket{W_{4,2}}$, which can be distilled to a Bell state
with a $1/3$ probability.

This protocol hence allows for Bell state generation from $4$ single photons with a probability of $50\%$ (without distillation) or $2/3$ (with distillation).

\subsection{Cost of classical control}
\label{sec:classical_control}
\begin{figure}
	\input{figures/mzi_subtraction.tex}
	\caption{
		\label{fig:mzi_subtraction}
    By replacing the BSG's four coupling beam splitters with these switchable couplers,
    we obtain switchable bleeding stages. 
    The open-circled couplers left and right are real, symmetric ones with
    reflectivity $\sin^2(\arcsin(\sqrt{r})/2)$. In the ``on'' state (phase set to $\pi$),
    the effect is that of a coupler with reflectivity $r$, in the ``off'' state it has unit transmissivity.
	}
\end{figure}
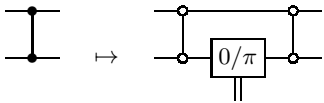
As always, after realizing the potential gains of a protocol, the ``no free lunch'' theorem strikes -- there's a non-negligible cost to achieving this improvement.
Repeated feed-forward comes with the price of repeated application of the potentially most lossy operation in this scheme: switching.
In order to recover the photon-subtracted state, subsequent bleeding stages have to be de-activated, or the state has to be switched in between stages to
some output ports. Either way, the signal state passes up to $S$ switches on its way to the output. The gain of scaling up the probability with more stages
seems to be counteracted by the accumulation of giant losses.
In fact, in practical implementations of linear optics circuits on chip, losses are considered to be the dominant source of imperfection -- measures to reduce
losses would require picking protocols with low switch depth to start with.

There are various ways to ensure a favorable scaling, depending on the characteristics of switches in the exact physical platform
used for implementation.
Good scaling can be obtained by making the four coupling beam splitters in $M$ switchable:
Each of the reflectivity-$r$ couplers in $\hat M(r)$ is replaced by an Mach-Zehnder-interferometer (MZI) made by two couplers
with reflectivity $\sin^2(\arcsin(\sqrt{r})/2)$ sandwiching an active $\pi/0$ phase shifter
(a controllable phase using e.g.\ thermo-optic or electro-optic effects) in one arm
(Fig.~\ref{fig:mzi_subtraction}), the effect of which is the ability to switch between $\hat M(r)$ and $\hat M(0)=\hat\id$.

\begin{figure}
  \input{figures/temporal_bleeding.tex}
  \caption{
    \label{fig:temporal_bleeding}
    A ``temporal'' bleeding circuit: measurements $M(r_k)$ ($M$ from Fig.~\ref{fig:bsg_standard} with coupling reflectivities switchable between $0$ and $r_k$)
    are applied repeatedly to $\fket{1}^{\otimes 4}$.
    The number of measured photons is added up and upon success (total of two photons detected), the switchable reflectivities in the remaining
    stages are turned off.
  }
\end{figure}
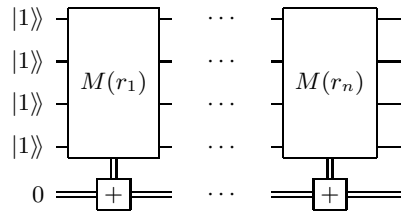

Combining multiple instances of the measurement circuit $M$ with stage-dependent reflectivities $r$ of the coupling beam splitters,
we can construct the full bleeding circuit shown in Fig.~\ref{fig:temporal_bleeding}.

In order to find the total loss rates due to switches in bleeding circuits with large number of stages,
let $\epsilon$ be the loss rate of an active phase shifter used in those switchable reflectivities.
Then, up to first order in $r$ and $\epsilon$,
the effective loss of this switch is $r\,\epsilon/4$, and its reflectivity is tunable between $0$ and $r (1-\epsilon/2)$.

Using constant reflectivities $r^\prime_k=1/S$ results in a constant switch depth, where the total loss due to
switches in the bleeding stages is $\epsilon/4$, independent of $S$.
For the previously discussed equal-spreading reflectivities from Eq.~\ref{eqn:equal_spread},
we can conclude (see Appendix~\ref{sec:bleeding_loss}) that the total loss rate is below a single switch loss $\epsilon$ for
bleeding schemes with up to $81$ stages.
It should be noted that those two choices,
however, have different limits for the success probability, so a small price may have to be paid to trade success probability
for lower loss.
Also, the toy model discussed here is by no means complete as it
only considers the dominant switch loss and neglects any losses due to other optical circuit elements,
such as the unavoidable delays required for any feed-forward based protocol.

Note that this is {\it one} possible arrangement and depending on the physical platform available, other setups with similar scaling are conceivable.

\subsection{Through space and time}
Instead of bleeding off small portions of the state's amplitude repeatedly, we can also ``spread'' it into all the possible measurement modes
($S+1$ per signal mode) in one initial step (see Appendix~\ref{sec:spatial_bleeding}).
Measurements are performed in switchable stages on $4$ modes each as before. Whenever the total of two photons have been detected in stage $k$, the
output state is still spread over $4\times(S+1-k)$ modes. Now, obtaining the $4$-mode output requires ``de-spreading'' $S+1-k$ modes into a single one.
This can be done probabilistically by passive interferometers or using switch-like reconfigurable networks.
Ultimately, combining switch networks in the detection and de-spreading stages allows to reduce ``switch depth'' and complexity of the circuits further.

In this ``spatial'' variant of bleeding switch requirements are different such that it might be preferable over the ``temporal'' version
given a specific hardware platform.

%% file: figures/bleeding_protocol.tex
\begin{center}\begin{tikzpicture}[node distance=2cm, auto]
  \tikzstyle{decision} = [diamond, draw, fill=blue!20, text width=4em, text badly centered, node distance=2.5cm, inner sep=0pt]
  \tikzstyle{block} = [rectangle, draw, fill=blue!10, text width=8em, text centered, rounded corners=1pt, minimum height=4em]
  \tikzstyle{line} = [draw, -latex']
  \tikzstyle{failure} = [draw, ellipse,fill=red!20, node distance=1cm, minimum height=2em]
  \tikzstyle{success} = [draw, ellipse,fill=green!20, node distance=1cm, minimum height=2em]

  \node [block] (init) {\vspace{-1em}\begin{eqnarray*}\ket{\psi}_{\mathrm{in}}&=&\fket{1}^{\otimes 4}\\N&=&0\end{eqnarray*}};
  \node [block, below of=init] (apply) {apply $\hat M$\\$n$: number of\\detected photons};
  \node [block, below of=apply] (increment) {$N=N+n$};
  \node [decision, below of=increment] (one) {$N \ge 2$?};
  \node [decision, below of=one] (two) {$N=2$?};
  \node [success, below of=two, node distance=2cm] (success) {success};
  \node [failure, left of=success, node distance=2cm] (failure) {failure};

  \path [line] (init) -- (apply);
  \path [line] (apply) -- (increment);
  \path [line] (increment) -- (one);
  \path [line] (one) --+(-2,0) node [near start] {no} |- (apply);
  \path [line] (one) -- node [near start] {yes} (two);
  \path [line] (two) -| node [near start] {no} (failure);
  \path [line] (two) -- node [near start] {yes} (success);
\end{tikzpicture}\end{center}

%% file: figures/bleeding_stages.tex
\begin{tikzpicture}
  \begin{axis}[
      width=1.05\columnwidth,
      height=.8\columnwidth,
      ylabel={$p_{\text{Bell}}$},
      xlabel={Number of stages $S$},
      ylabel style={at={(.05,.5)}},
      xmin=0, xmax=10.2,
      ymin=0, ymax=0.5,
      xtick={0,2,4,6,8,10},
      ytick={0.0,0.1,0.2,0.3,0.4,0.5},
      ymajorgrids=true,
      grid style=dashed,
      legend style={
        draw=none,
        fill=none,
        at={(1.,.2)},
        column sep=8pt
      },
      legend cell align={left},
  ]
  \addplot[
      only marks,
      mark size=1.5pt,
      draw=none,
      color={blue!80},
      ]
      coordinates {
        (1,0.1875)
        (2,0.27051)
        (3,0.318029)
        (4,0.348983)
        (5,0.370812)
        (6,0.387058)
        (7,0.399634)
        (8,0.409665)
        (9,0.417856)
        (10,0.424673)
        (11,0.430437)
        (12,0.435376)
        (13,0.439655)
        (14,0.443399)
        (15,0.446703)
        (16,0.449641)
        (17,0.45227)
        (18,0.454636)
        (19,0.456778)
        (20,0.458726)
      };
  \addplot[
      mark=square*,
      only marks,
      mark size=1.5pt,
      draw=none,
      color={orange!90!black},
      ]
      coordinates {
        (1,0.1875)
        (2,0.259259)
        (3,0.304688)
        (4,0.336)
        (5,0.358796)
        (6,0.376093)
        (7,0.389648)
        (8,0.400549)
        (9,0.4095)
        (10,0.41698)
        (11,0.423322)
        (12,0.428766)
        (13,0.433491)
        (14,0.43763)
        (15,0.441284)
        (16,0.444535)
        (17,0.447445)
        (18,0.450066)
        (19,0.452438)
        (20,0.454595)
      };
    \legend{optimal, equal spread}
  \end{axis}
\end{tikzpicture}

%% file: figures/mzi_subtraction.tex
  \centerline{\Qcircuit @C=1em @R=1em {
    & \control \qw & \qw & & & & & \controlo \qw & \qw & \controlo \qw & \qw   \\
    & \ctrl{-1} & \qw & & \mapsto & & & \ctrlo{-1} & \gate{0 / \pi} & \ctrlo{-1} & \qw \\
    & & & & & & & & \cwx & & &
  }\vspace{1em}}

%% file: figures/temporal_bleeding.tex
\newcommand{\cgate}[1]{*+<.6em>{#1} \POS ="i","i"+UR;"i"+UL **\dir{-};"i"+DL **\dir{-};"i"+DR **\dir{-};"i"+UR **\dir{-},"i" \cw}
\centerline{\Qcircuit @C=.5em @R=.9em {
  \lstick{\fket{1}} & \multigate{3}{M(r_1)} & \qw &
    \push{\quad\cdots\quad} & &
    \multigate{3}{M(r_n)} & \qw & \qw \\
  \lstick{\fket{1}} & \ghost{M(r_1)} & \qw &
    \push{\cdots} & &
    \ghost{M(r_n)} & \qw & \qw \\
  \lstick{\fket{1}} & \ghost{M(r_1)} & \qw &
    \push{\cdots} & &
    \ghost{M(r_n)} & \qw & \qw \\
  \lstick{\fket{1}} & \ghost{M(r_1)} & \qw &
    \push{\cdots} & &
    \ghost{M(r_n)} & \qw & \qw \\
  \lstick{0} & \cgate{+}\cwx & \cw &
    \push{\cdots} & &
    \cgate{+}\controlo\cwx\cw & \cw & \cw
}\vspace{1em}}

%% file: sections/primates.tex
In this section we are going to present a scheme to generate $n$-photon GHZ states from $2n$ single photons. The scheme allows intermediate states generated during the creation of the GHZ states to be multiplexed and can also utilize the deterministic photon subtraction method, presented in Sec.~\ref{sec:bleeding}, to increase the state generation probability.

\subsection{GHZ generation}
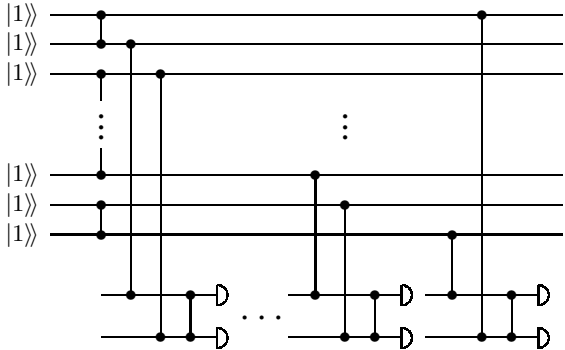
\begin{figure}
  \input{figures/ghz_generator.tex}
  \caption{
    A circuit to produce $n$-GHZ states from $2n$ single photons. For $n=2$ this coincides with the standard
    BSG circuit discussed above with re-arranged beam splitters, up to a $H\otimes H$ correction on the output qubits.
  }
  \label{fig:ghz_generator}
\end{figure}
Before presenting the new scheme, we want to remind the reader of how $n$-GHZ states can be produced by linear optical means.
The circuits we consider here (see Fig.~\ref{fig:ghz_generator}) are based on those presented in~\cite{Varnava2008} for $n=3$ and generalized to $n>3$ qubits in~\cite{Mercedes2015}
\footnote{Similar to the BSG where the projective measurement of the ancillae turned out to be a Bell state measurement, the ancilla projection in a GHZ generator
is a GHZ state discrimination circuit which can be used as a dual-rail $n$-partite fusion.}.
Table~\ref{tab:n_ghz_succ_probs} shows the probability of generating an $n$-qubit GHZ state from $2n$ single photons.

In the case of $n=2$ we can make some comparisons to the Bell state generation schemes presented earlier. 
Without bleeding, the success probability for this scheme of $1/8$ is equal to the one from standard Bell state generation when only detection outcomes which leave the state in a set of pre-determined dual rails are accepted, i.e.\ using only four of the six detection outcomes described in Sec.~\ref{sec:bsg_corrections}.
This alludes to the fact that the beam splitters can be rearranged into the same layout used in the earlier section, with the addition of beam splitters acting on the vacuum inputs (with no effect) and those implementing transfer matrices $H^{(2)}\oplus H^{(2)}$ on the output modes (which perform the qubit-level unitary $H\otimes H$ on the output states, the effect of which is a permutation of the generated Bell states).

\subsection{Non-qubit intermediate states}
The scheme shown in Fig.~\ref{fig:ghz_generator} can be decomposed into smaller steps which correspond to fusing states of the form
\begin{equation}
  \begin{split}
    \ket{\pi_n^{\lambda, \pm}} =& \sqrt{\lambda}
      \frac{
        \fket{2}\fket{01}^{\otimes{n-1}}\fket{0} \pm \fket{0}\fket{10}^{\otimes{n-1}}\fket{2}
      }
      {
        \sqrt{2}
      }
    \\
    &+ \sqrt{1-\lambda} \fket{0}\fket{\zeta}\fket{0},
  \end{split}
\end{equation}
which we refer to as $n$-primates. $n$-primates are a set of states for $\lambda \in [0, 1]$ and arbitrary normalized $\fket{\zeta}$. The specifics of $\fket{\zeta}$ play no role in the scheme.

\begin{table}
  \begin{tabular}{ll}\toprule
    Bleeding & Success probability \\ \midrule
    No       & $1/2^{2n-1}$ \\
    Yes      & $1/2^{n-1}$ \\ \bottomrule
  \end{tabular}
  \caption{Success probabilities for generating $n$-qubit GHZ states from $2n$ single photons. The probability presented with bleeding is in the limit of a large number of stages.}
  \label{tab:n_ghz_succ_probs}
\end{table}

\begin{figure}
  \input{figures/primate_elements.tex}
  \caption{
    Elementary circuits used to generate GHZ states in Fig.~\ref{fig:ghz_generator}. The first circuit shows the generation of a $1$-primate from two single photons.
    The second circuit shows the circuit used to fuse primates together (see Appendix~\ref{sec:primate_expression} for explicit measurement operators for $F(t)$).
    The transmissivity of the two labeled beam splitters can be any equal value (emphasized by using open circles instead of 50:50 beam splitters)
    and the unlabeled beam splitter is a 50:50 beam splitter. The fusion operation succeeds when one photon is detected at either detector.
  }
  \label{fig:primate_elements}
\end{figure}
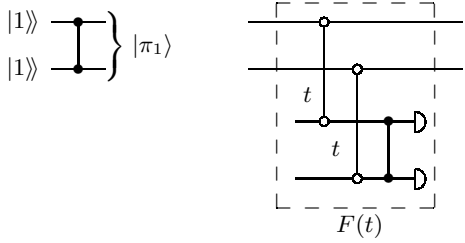

\begin{figure}
  \input{figures/fusing_primates.tex}
  \caption{Fusions between primates and subsequently conversion into a GHZ state.}
  \label{fig:fusing_primates}
\end{figure}
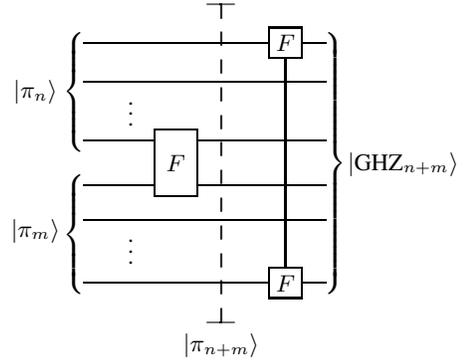

The circuit used to fuse primates is shown in Fig.~\ref{fig:primate_elements} together with the circuit required to generate a $1$-primate deterministically from two single photons. From a pair of $n_1$- and $n_2$-primates, the bigger $n$-primate (with $n=n_1+n_2$) can be generated by applying the fusion operation on an outermost mode of each of the input states as shown in Fig.~\ref{fig:fusing_primates}. An $n$-primate can be turned into an $n$-qubit GHZ state by applying the same fusion operation on the outermost modes of a single state. The fusion operation succeeds when one photon is detected at either detector. If no photons are detected, the fusion operation can be retried until a single photon is detected, similar to the bleeding schemed presented in Sec.~\ref{sec:bleeding}. Performed in this way, the fusion can succeed with the probability that there is at least one photon in the modes undergoing fusion. In Appendix~\ref{sec:primate_expression} we compute the success probabilities of these fusion operations in terms of $\lambda$. It is important to note that the transmissivity of the beam splitters used in the fusion determines the $\lambda$ of the resultant primate. Specifically, a lower transmissivity will yield a higher $\lambda$ which in turn will have a higher success probability in subsequent fusions. One should also note that the success probabilities of these fusions are not independent from one another.

\subsection{Bleeding}
It is interesting also to compare the success probabilities for Bell state generation using the bleeding strategy described in this section to that of the previous sections. The success probability in the limit of a large number of stages is equal at $1/2$ as well, but the scheme presented in this section will leave the Bell-state in a predetermined set of modes, unlike the strategy presented earlier. The difference is that in the previous section all four modes are bled together until four photons are detected, whereas here pairs of modes are bled independently until one photon from each pair is detected.

\subsection{Multiplexing}
\label{sec:primates_multiplexing}
We have been deliberate in presenting the scheme in a modular way such that it is clear that intermediate stages can be multiplexed to discard intermediate states involved in failed fusions. Multiplexing can be used to increase the resource utilization in between two subsequent fusions of different primates. Nothing can be gained by multiplexing in between the final two fusions as they occur on the same pair of primates. It is worthwile noting that the concept of multiplexing here is the same as for qubit states and the same switching networks can be employed.

To provide an example of the possible resource savings offered by this intermediate multiplexing,
we are going to use the notion of \emph{perfectly resource efficient multiplexing}\footnote{An extensive review of switching schemes can be found in~\cite{PsiQ_SN}.}:
This describes the process where from a reservoir of $N$ probabilistic resources (each being present with probability $p_0$) not a single, but $M > 1$ copies are selected.
In a situation where e.g.\ a fusion operation is to be performed on multiple pairs in parallel, all those fusions can be fed by the same MUX.
In the limit $N\rightarrow\infty$, $Np_0$ resources are present. By choosing $M=Np_0$, \emph{all} available resources can be utilized downstream.

We now compare the average number of single photons required to produce a 4-qubit GHZ state with and without using perfectly resource efficient multiplexing in between successive fusions.
This allows us to compute the number of single photons needed to obtain a particular state recursively:
With $N$ being the number of single photons required to generate the input to a fusion operation, $p_0$ the probability of the fusion succeeding and $N^\prime=N/p_0$ is the number of single photons required to generate the post fusion state.

We consider four ways of generating a four GHZ state none of which utilize bleeding:
\begin{enumerate}
  \item Generation from $8$ single photons without intermediate multiplexing ($p=1/128$).
  \item Generation from $10$ single photons, $6$ of which are converted to a 3-GHZ state ($p=1/32$), and $4$ of which into a Bell state ($p=1/8$) on a pre-determined set of modes.
    GHZ states and Bell pairs are then multiplexed before being fused into a 4-GHZ state with a Type-I fusion ($p=1/2$).
  \item Generation from $12$ single photons, via $3$ Bell pairs. Those are fused using a Type-I fusion in two steps, first to obtain a 3-GHZ and a Bell pair, and finally a 4-GHZ.
  \item  Generation from $8$ single photons via 2-primates which are then multiplexed into the final fusions which form the GHZ state. The optimal transmissivity used to generate the 2-primates is $t=\sqrt{2}-1$. The value of $t$ is chosen to maximize the success probability of the entire protocol, it is not the transmissivity which maximizes the success probability for that stage. In the final fusion stage the optimal transmissivity is always $t=1/2$ as it maximizes success probability and there is no post fusion $\lambda$ to consider. The full calculation for the photon consumption of this scheme is shown in Appendix~\ref{sec:primate_expression}.
\end{enumerate}

The average number of single photons required to generate a 4-GHZ state with each of these methods is presented in Table~\ref{tab:four_ghz_resources}. As can be seen in the table, the primates scheme offers significant resource savings. We note that the states produced by these methods will have other distinguishing features when imperfections are taken into account, which must be considered to assess their true efficacies.

For multiplexing schemes that are not perfectly resource efficient, reductions in the required resources can sometimes be obtained by prioritizing states with higher $\lambda$.
The variations in $\lambda$ can be introduced by using a set of different fusion transmissivities at a particular fusion stage but will also be caused by bleeding due to successes at different stages.
For example, consider a situation where a multiplexing scheme is used to increase the probability that an $n$-primate is present.
The multiplexing scheme selects one post-fusion state from two preceding fusion operations, $\ket{\pi^{\lambda_1}}$ and $\ket{\pi^{\lambda_2}}$.
Those are present when the first or second fusions succeed, respectively.
If the scheme always selects the post-fusion state $\ket{\pi^{\lambda_1}}$, then
this state should be produced with higher $\lambda_1$.
This can be achieved by altering the transmissivity of the fusions, at the cost of reducing the probability of producing this state, but in turn
increasing the resource efficiency of the entire scheme. 

\begin{table}[]
  \setlength{\tabcolsep}{0.5em}
  \begin{tabular}{@{}ll}\toprule
    Scheme                                                                             & Num. Single Photons           \\ \midrule
    $\fket{1}^{\otimes 8} \rightarrow \ket{\text{GHZ}_4}$                                     & 1024                          \\
    $\fket{1}^{\otimes 10} \rightarrow \ket{\text{GHZ}_2}\ket{\text{GHZ}_3} \rightarrow \ket{\text{GHZ}_4}$ & 448                           \\
    $\fket{1}^{\otimes 12} \rightarrow \ket{\text{GHZ}_2}^{\otimes 3}$ & 320 \\
      $\qquad\rightarrow \ket{\text{GHZ}_3}\ket{\text{GHZ}_2} \rightarrow \ket{\text{GHZ}_4}$ & \\
    $\fket{1}^{\otimes 8} \rightarrow \ket{\pi_2}^{\otimes 2} \rightarrow \ket{\text{GHZ}_4}$  & $128(1+\sqrt{2}) \approx 309$ \\ \bottomrule
  \end{tabular}
  \caption{
    Average number of photons required to generate a four qubit GHZ state.
    While the costs of the third and fourth schemes do not differ that much in this comparison, it should be noted that
    with realistic (finitely-sized) multiplexing, the scheme with the lower resource consumption per attempt (primates with only $8$ photons) will be advantageous.
    Also, a lower number of input per attempt reduces the errors (due to imperfect sources or misalignments in the network itself)
    accumulated in the output state.
  }
  \label{tab:four_ghz_resources}
\end{table}

Finally we want to mention that one main driver for GHZ state generation is the application of those states in
fusion-based schemes for building larger resource states for quantum computing.
We note that it is not necessary to fully form a GHZ state before performing subsequent qubit operations such as Type-I and Type-II fusions.
As an example, a Type-I fusion could be performed between modes from two primates provided that none of the modes entering the Type-I fusion are the outer modes of either of the primates.
With the remaining primate fusions occurring after the Type-I fusion, the resulting state (upon success) will be the same as if Type-I had occurred between two GHZ states.

%% file: figures/ghz_generator.tex
\centerline{\Qcircuit @C=1em @R=1em {
  \lstick{\fket{1}} & \qw & \control \qw & \qw & \qw & \qw & \qw & \qw & \qw & \qw & \qw & \qw & \qw & \qw & \qw & \qw & \control\qw & \qw & \qw & \qw \\
  \lstick{\fket{1}} & \qw & \control \qw\qwx & \control \qw & \qw & \qw & \qw & \qw & \qw & \qw & \qw & \qw & \qw & \qw & \qw & \qw & \qw & \qw & \qw & \qw \\
  \lstick{\fket{1}} & \qw & \control \qw\qwx[1] & \qw & \control \qw & \qw & \qw & \qw &\qw & \qw & \qw & \qw & \qw & \qw & \qw & \qw & \qw & \qw & \qw & \qw \\
  &&&\\
  &&\parbox{1em}{\Large$\vdots$\\\vspace{.3em}}&&&&&&&&&\parbox{1em}{\Large$\vdots$\\\vspace{.3em}}&\\
  &&&\\
  \lstick{\fket{1}} & \qw & \control \qw\qwx & \qw & \qw & \qw & \qw & \qw & \qw & \qw & \control \qw & \qw & \qw & \qw & \qw & \qw & \qw & \qw & \qw & \qw \\
  \lstick{\fket{1}} & \qw & \control \qw & \qw & \qw & \qw & \qw & \qw & \qw & \qw & \qw & \control \qw & \qw & \qw & \qw & \qw & \qw & \qw & \qw & \qw \\
  \lstick{\fket{1}} & \qw & \control \qw\qwx & \qw & \qw & \qw & \qw & \qw & \qw & \qw & \qw & \qw & \qw & \qw & \qw & \control \qw & \qw & \qw & \qw & \qw \\
  & & & & & & & & & & & \\
  & & & \ctrl{-9} & \qw & \control\qw     & \detector & & & & \ctrl{-4} & \qw & \control\qw     & \detector & & \ctrl{-2}  & \qw & \control\qw     & \detector & \\
  & & & \qw & \ctrl{-9} & \control\qw\qwx & \detector & & \parbox{1cm}{\Large$\cdots$\\\vspace{1em}} & & \qw & \ctrl{-4} & \control\qw\qwx & \detector & & \qw & \ctrl{-11} & \control\qw\qwx & \detector &
}\vspace{2em}}

%% file: figures/primate_elements.tex
\centerline{\Qcircuit @C=1em @R=1.6em {
  &&\lstick{\fket{1}}&\ctrl{1}   &\qw &&\raisebox{-2.5em}{$\ket{\pi_1}$}&&&&&\qw&\qw                                     &\ctrlo{2}                               &\qw        &\qw        &\qw      &\qw&\qw\\
  &&\lstick{\fket{1}}&\control\qw&\qw &&                               &&&&&\qw&\qw\rule{1em}{0em}\raisebox{-2.5em}{$t$}&\qw                                    &\ctrlo{2}   &\qw        &\qw      &\qw&\qw\\
  &&                 &           &    &&                               &&&&&   &                                        &\controlo\qw                            &\qw        &\ctrl{1}   &\detector&\\
  &&                 &           &    &&                               &&&&&   &                                        &\qw\rule{1em}{0em}\raisebox{2.5em}{$t$}&\controlo\qw&\control\qw&\detector&\\
  &&                 &           &    &&                               &&&&&   &                                        & \rule{3em}{0em}F(t)                   &\\
  {\gategroup{1}{5}{2}{5}{0.8em}{\}}}
  {\gategroup{1}{13}{4}{17}{1.6em}{--}}}
}

%% file: figures/fusing_primates.tex
\centerline{\Qcircuit @C=1em @R=1.0em {
                                 &&&&&&& \\
                                 &&&\qw& \qw  &\qw& \qw              &\qw& \sgate{F}{7} &\qw \\
  \raisebox{-1.3em}{$\ket{\pi_n}$}&&&\qw& \qw  &\qw                  & \qw              &\qw& \qw&\qw\\
                                 &&&   &\vdots&                     &                  &   &    \\
                                 &&&\qw& \qw  &\multigate{1}{F}     & \qw              &\qw&\qw&\qw&&&{{\raisebox{-2.5em}{\hspace{0.5em}$\ket{\text{GHZ}_{n+m}}$}}}\\
                                 &&&\qw& \qw  & \ghost{F}           & \qw              &\qw&\qw&\qw  \\
  \raisebox{-1.4em}{$\ket{\pi_m}$}&&&\qw& \qw  & \qw                 & \qw              &\qw&\qw&\qw \\
                                 &&&   &\vdots&                     &                  &   & \\
                                 &&&\qw& \qw  & \qw                 & \qw              &\qw&  \gate{F}&\qw  \\
                                 &&&&&&& \\
                                 &&&   &      &                     &{\ket{\pi_{n+m}}}&\\
  {\gategroup{2}{3}{5}{3}{0.8em}{\{}}
  {\gategroup{6}{3}{9}{3}{0.8em}{\{}}
  {\gategroup{1}{7}{10}{7}{0em}{--}
  {\gategroup{2}{10}{9}{10}{0.8em}{\}}}}
  }
}

%% file: sections/boosting.tex
\subsection{Background}
Fusion~\cite{BR04} was initially introduced as a linear optical operation which allowed to probabilistically ``stick together'' pieces of cluster state~\cite{RB01}. For the purposes of more general fault-tolerant photonic quantum computing, wherein we are not creating cluster states per se~\cite{PsiQ_FBQC}, it is useful to define fusion in terms of POVMs that have (subsets of) outcomes which reveal nontrivial stabilizer-valued correlations (e.g.\ Pauli $XX$ and $ZZ$ outcome bit values) between dual-rail qubits .

From the perspective of underlying photonic protocols, however, this latter way of thinking can potentially be too limiting. Therefore, we consider a more Hilbert-space focused way of thinking about fusion. For pedagogical clarity, imagine we single out two qubits, each maximally entangled with distinct subsets of qubits labeled $A,B$:
\begin{equation} 
  \frac{1}{\sqrt{2}}\left(\ket{A_0}\!\ket{0}+\ket{A_1}\!\ket{1}\right)\otimes \frac{1}{\sqrt{2}}\left(\ket{B_0}\!\ket{0}+\ket{B_1}\!\ket{1}\right) 
  \label{eqn:AcrossB}
\end{equation}
 
A successful fusion is a two-qubit measurement outcome that leaves the $A,B$ qubits entangled in a desirable way. An example would be a non-destructive two-qubit parity measurement given by projectors, 
\begin{equation}
  \Pi_E=\ket{00}\!\bra{00}+\ket{11}\!\bra{11},\,\Pi_O=\ket{01}\!\bra{01}+\ket{10}\!\bra{10} \nonumber
\end{equation}
which will collapse the whole system to a state of the form
$\ket{A_0B_0}\!\ket{00}+\ket{A_1B_1}\!\ket{11}$ or $\ket{A_0B_1}\!\ket{01}+\ket{A_1B_0}\!\ket{10}$. 

In photonics the simplest measurements are fully destructive, but as shown in~\cite{BR04} --
wherein the operation is termed \emph{Type-II fusion} -- destroying the two measured qubits and obtaining entangling
outcomes of the form $\ket{A_0B_0}\pm\ket{A_1B_1}$ or $\ket{A_0B_1}\pm\ket{A_1B_0}$
(with any finite probability) is sufficient to achieve full quantum computational
universality, with this joint measurement as the only non-trivial (i.e.\ 2-qubit) primitive.  

Type-II fusion can be achieved by any measurement that has at least some outcomes which project onto Bell states. There is a long history of studying how well a linear optical measurement containing outcomes that project into the four standard dual rail Bell basis can be performed. It was shown early on~\cite{CL01} that such Bell state measurements (BSMs) will achieve at most a $50\%$ efficiency in distinguishing the four Bell states if only vacuum ancillae are allowed. Significant improvements~\cite{Grice11,EL14} can be made by using ancillary photonic states to \emph{boost} the efficiency past the $50\%$ threshold. In Sec.~\ref{sec:fusion_vs_bsm} we present a scheme for boosted Type-II fusion that is not a BSM, and that has a higher success probability than the scheme of~\cite{EL14} for the same ancillary resources.

Type-I fusion involves destruction of only one qubit. Using the example given above, a successful Type-I fusion outcome would be one that projected the qubits of Eq.~(\ref{eqn:AcrossB}) into a state of the form $\ket{A_0B_0}\!\ket{0}+\ket{A_1B_1}\!\ket{1}$, or $\ket{A_0B_1}\!\ket{0}+\ket{A_1B_0}\!\ket{1}$ (or similar states with a Pauli operation on the qubit that has survived the fusion). Type-I fusion typically facilitates considerable resource savings, not only because it destroys one fewer qubit, but because the qubit that remains is entangled in such a way that it is a useful substrate onto which larger entangled states can be fused. It is worth noting that the advantage of destroying only one qubit during Type-I fusion comes at the cost of losing the loss tolerance that Type-II fusion offers, a benefit stemming from the fact that the latter exploits the error-detecting properties of dual-rail encoding by counting the photons in all rails belonging to the qubits it acts on.

In the next section we present a scheme for boosting Type-I fusion, a primitive previously unknown to be boostable, and then discuss one of its implications, namely that it essentially resolves in the affirmative the open question of whether there exists a ``ballistic with Bell pairs'' percolation scheme~\cite{KRE07,GSBR15,Rudolph17}
for generation of large cluster states.

\subsection{Boosted Type-I fusion}
\label{sec:boosted_type1}
\subsubsection{Unboosted Type-I fusion}
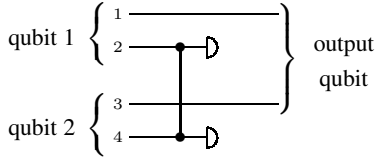
\begin{figure}
  \input{figures/unboostedTypeI.tex}
  \caption{
    A standard Type-I fusion: One mode each from a pair of input qubits
    is subjected to a collective measurement.
    The remaining two modes are considered the output qubit in case of success (projection of the measured modes onto $\fket{10}\pm\fket{01}$).
    \label{fig:unboostedTypeI}
  }
\end{figure}

Unboosted Type-I fusion is performed by the circuit in Fig.~\ref{fig:unboostedTypeI}.

To understand why it works, we expand the initial state in Eq.~(\ref{eqn:AcrossB}), explicitly writing the qubit states as dual rail encoded photonic states:
\begin{align}
  \frac{1}{2}\big(&\ket{A_0}\!\ket{B_0}\!\fket{10}\!\fket{10}+\ket{A_1}\!\ket{B_1}\!\fket{01}\!\fket{01} \nonumber\\
  + &\ket{A_0}\!\ket{B_1}\!\fket{10}\!\fket{01}+\ket{A_1}\!\ket{B_0}\!\fket{01}\!\fket{10}\big) \nonumber
\end{align}
Modes $2$ and $4$ undergo a 2-Hadamard measurement, which projects onto $\fket{10}\pm\fket{01}$, $\fket{00}$ and $\fket{11}$. We see the $\fket{10}\pm\fket{01}$ outcomes do an unbiased projection into the last two terms of the above equation, leaving a fused state $\ket{A_0B_1}\!\fket{10}\pm\ket{A_1B_0}\!\fket{01}$. Note the the two modes $1$,$3$ of the fused qubit each originate from a different original qubit. The $\fket{00}$ and $\fket{11}$ outcomes are failures that project into unentangled states of the form $\ket{A_0}\!\ket{B_0}\!\fket{11}$ or $\ket{A_1}\!\ket{B_1}\!\fket{00}$. Success occurs with probability $1/2$.

\subsubsection{Boosting Type-I fusion to success probability $3/4$}
The key to boosting the success probability of the Type-I fusion is to find a suitable ancilla to render indistinguishable the two distinguishable failure outcomes of the unboosted protocol. However, the states $\fket{00}$ and $\fket{11}$ have different numbers of photons. While it is possible to use ancillary photonic states such as $\fket{00}+\fket{11}$ to do measurements with some outcomes that render these states indistinguishable, doing so creates overall states of indefinite total photon number. This is typically undesirable because detection of photon loss is much easier if we know the total number of photons we are expecting.

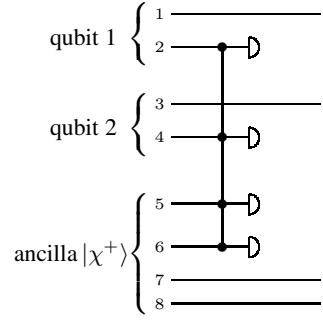
\begin{figure}
  \input{figures/boosted_type_1.tex}
  \caption{
    A ``boosted'' Type-1 fusion: The collective measurement includes also two modes from
    a four-mode ancilla state which can increase the rate of successful outcomes from $1/2$ to $3/4$.
    \label{fig:boosted_type_1}
  }
\end{figure}

The way to boost without generating states of indefinite photon number is to use an ancillary photonic state of the form $\ket{\chi^+}=2^{-1/2}(\fket{11}\!\fket{00}+\fket{00}\!\fket{11})$. Note that this ancilla is simply a dual rail Bell state with modes permuted. Boosted Type-I fusion can then be achieved via the photonic circuit shown in Fig.~\ref{fig:boosted_type_1}.

The Kraus operators describing the effects of the circuit in Fig.~\ref{fig:boosted_type_1} are $\sqrt{w_j}K_j$ where
\begin{align}
  K_{1\pm}&=\fket{1011}\!\bra{01}\pm\fket{0111}\!\bra{10} \nonumber\\
  K_{2\pm}&=\fket{1000}\!\bra{01}\pm\fket{0100}\!\bra{10} \nonumber\\
  K_{3\pm}&=\fket{1100}\!\bra{00}\pm\fket{0011}\!\bra{11} \nonumber\\
  K_{4}&=\fket{0000}\!\bra{11} \nonumber\\
  K_{5}&=\fket{1111}\!\bra{00} \nonumber,
\end{align}and 
\begin{equation}
  w_{1\pm}=w_{2\pm}=w_{3\pm}=\frac{1}{4}, \,\,w_4=w_5=\frac{1}{2}. \nonumber
\end{equation}

Each of these operators corresponds to a subset of the 42 possible successful detection patterns of this protocol.
The circuit succeeds whenever we obtain any of the first six outcomes, noting that $K_{1\pm}^\dagger K_{1\pm}=K_{2\pm}^\dagger K_{2\pm}=\Pi_O$, $K_{3\pm}^\dagger K_{3\pm}=\Pi_E$.

The input states for this measurement are two dual-rail qubits, however the output states are given in the Fock basis because not all outputs are qubits. As per the figure, the four output modes are  $1$,$3$,$7$,$8$. If outcomes $K_{1\pm}$ or $K_{2\pm}$ are obtained, then modes $1$ and $3$ contain an output dual-rail qubit, while modes $7$,$8$ factorize out (into non-qubit states $\fket{00}$ or $\fket{11}$). If outcomes $K_{3\pm}$ are obtained then we can combine modes $1$,$7$ into one dual rail qubit, and modes $3$,$8$ into a different one. We see that in this latter case we have performed Type-I fusion and actually left two entangled qubits in the state. To summarize, with probability $1/2$ we fuse to $\ket{A_0B_1}\!\ket{0}+\ket{A_1B_0}\!\ket{1}$ and with probability 1/4 we fuse to $\ket{A_0B_0}\!\ket{00}+\ket{A_1B_1}\!\ket{11}$. The total success probability of $3/4$ matches the success probability for Type-II fusion based on the boosted BSM of~\cite{Grice11}. Both schemes use a single Bell pair as an ancilla.

To understand why this circuit works note that the 4-Hadamard measurement explained in Sec.~\ref{sec:bsg}
is performed on modes $2$,$4$,$5$,$6$. In the unboosted case failure occurred when modes $2$,$4$ were in $\fket{00}$
or $\fket{11}$. However, with the ancilla now part of the measurement, there is a part of the
total input state where these two terms combine with the two measured ancilla modes into two
photon terms of the form $\fket{00}\!\fket{11}$ and $\fket{11}\!\fket{00}$. Looking at the
effective POVM for the 4-Hadmard we see that all of the measurement operators involving
two input photons project unbiasedly onto $\fket{00}\!\fket{11}\pm\fket{11}\!\fket{00}$.
This becomes a successful fusion (operators $K_{3\pm}$). Parts of the input state with
$\fket{00}\!\fket{00}$ or $\fket{11}\!\fket{11}$ in modes $2$,$4$,$5$,$6$ will lead to failure.
Thus we have converted some fraction of previously unsuccessful outcomes into successful ones.

It is also important to confirm that bringing in an ancilla and doing a different interferometer has not turned the previously successful outcomes into failures. In the unboosted case modes $2$,$4$ contained $\fket{10}$ or $\fket{01}$ when the protocol succeeded. When combined with the two measured modes of the ancilla, this part of the input state involves either one or three photons. Consider the case where the input is of the form $\fket{10}\!\fket{11}$ or $\fket{01}\!\fket{11}$. Once again looking at the measurement operators for the 4-Hadamard in Appendix~\ref{sec:h4_projection} we see that all of the three-photon outcomes project unbiasedly onto $\fket{10}\!\fket{11}\pm\fket{01}\!\fket{11}$ ($K_{1\pm}$), thus leading to the same success outcomes as for unboosted fusion. A similar reasoning can be applied in the case where no ancilla photons are present in the measured modes, giving rise to $K_{2\pm}$.
The unboosted protocol succeeded when only one photon was detected. The boosted protocol succeeds when one ($K_{2\pm}$) or three photons ($K_{1\pm}$) are measured (corresponding to the same outcomes as the unboosted case) or when two photons (one from the state, one ancilla) are detected ($K_{3\pm}$). In the latter case, we entangle an extra ancilla photon with the state, which can later be used as an anchor for future fusion operations.

In the scheme presented, the final evolution, for either success or failure, can be understood purely as a (dual-rail qubit) stabilizer CP map. However, at intermediate stages we necessarily made use of the ``single rail W-type'' outcomes from the 4-Hadamard effective POVM, in addition to the ``single rail stabilizer state'' outcomes. The latter, in the case of Bell state generation, \emph{are} actually the most interesting/useful. Boosting Type-I fusion is therefore a cautionary tale: when targeting fault-tolerant computation the general obsession with stabilizers does make it tempting to prematurely focus attention on (easily understood) parts of circuits that are ``stabilizer like''.

If we wish to ``double boost'' Type-I fusion we need a way of rendering the states $\fket{0000}$ and $\fket{1111}$ indistinguishable. This can be done using a 4-GHZ state as an ancilla with a success probability of $7/8$. This highlights a general principle for boosting that is somewhat algorithmic, namely look at the failure modes of an operation, and try to ``lift'' them into successes via bringing in an ancilla with similar structure. The subtlety is that the process of doing so needs to preserve the successful outcomes that have already been attained.

\subsubsection{Implications for ballistic generation of cluster states}
Early approaches to photonic quantum computing used either ``gate teleportation + post-selection''~\cite{KLM01} or ``repeat until success''~\cite{BR04}, or a combination of both~\cite{Nielsen04}, as the underlying mechanisms whereby probabilistic photonic protocols were rendered effectively deterministic so as to create extensive entanglement.

Such architectures make heavy use of \emph{active} photonic elements, such as switches and quantum memories. Active elements are much more difficult to build (and more lossy) than passive elements, such as interferometers. In~\cite{KRE07} it was shown that starting with small photonic cluster states, a single non-adaptive step of applying passive fusion gates simultaneously could be used to generate extensive entanglement. This scheme was improved upon in various ways~\cite{GSBR15,PsiQ_FBQC,pant2019percolation,morley2017physical}, however to date the smallest size of the initial cluster state for which such a ``ballistic'' scheme was known to be possible was a 3-GHZ state. 

Using boosted Type-I fusion (which requires only a Bell state as an ancilla) we can perform fusion of three qubits with probability $(3/4)^2\approx 56\%$. That is, we can fuse qubits $1$ and $2$ and then if successful fuse in qubit 3. There are trivalent graphs (such as a tree) with site percolation thresholds lower than this, though no simple lattices with such low threshold are known to us. Regardless, it is not difficult to see that by placing Bell states on the edges of such graphs (e.g.\ graphs built from suitably interlinked trees) and performing three-way boosted Type-I fusions we can create, via a single percolation event, extensive entanglement that can be re-normalized into a computationally universal cluster state. 

The caveat to this observation is that it does not fully play by the rules of a completely ballistic protocol as defined in~\cite{KRE07}. This is because in the boosted Type-I some of the outcomes require a permutation of modes in order to define where the output qubit is. Abstractly one may think this can be achieved via passive re-labeling, however since we are concatenating two Type-I's an active element must be used to select which output modes from the first boosted fusion are sent into the second. This requires one step of (local) adaptivity.

\subsection{Type-II fusion vs. Bell state measurement}
\label{sec:fusion_vs_bsm}
For the purposes of fusion as a quantum computational primitive it is frequently the case that fusion into a more general state of the form $\ket{A_0B_0}+e^{i\phi}\ket{A_1B_1}$ is sufficient, as long as $\phi$ is known. This is because the structure of the $\ket{A_i}$ or $\ket{B_i}$ states often allows for correction of the phase $\phi$ via an easily performed linear optical unitary. A measurement outcome described by a Kraus operator of the form $\sqrt{w}\left(\bra{00}+e^{i\phi}\bra{11}\right)/\sqrt{2}$, for example, would implement this fusion on the qubits of Eq.~(\ref{eqn:AcrossB}) with probability $w/4$.

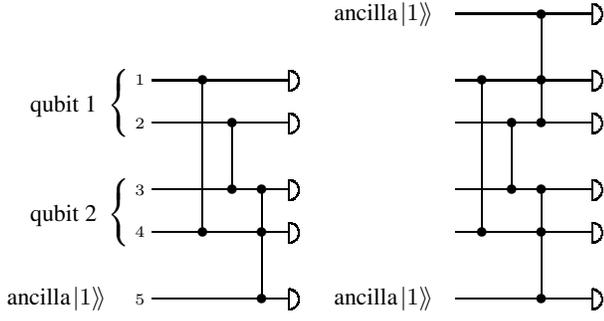
\begin{figure}
  \input{figures/boostedtype2.tex}
  \caption{
    ``Boosted'' Type-II fusions: In these minimally boosted versions,
    single-mode ancilla states are used to increase the success probability from
    $1/2$ to $7/12$ (left) and $2/3$ (right). The single-mode ancillae are
    interfered with two of the signal modes each via $3\times 3$ multiports defined by
    DFT transfer matrices (simply termed 3-DFT).
    Also, the ``success'' outcomes are no longer those of a Bell state measurement.
    \label{fig:boostedtype2}
  }
\end{figure}
To understand why this observation can be useful, consider the interferometer of Fig.~\ref{fig:boostedtype2} (left). We are interested in the measurement performed when the four input modes indicated are restricted to being dual rail qubit states. A simple, but tedious, calculation reveals that the effective measurement operators on the two input qubits take the form $\sqrt{w_i}\bra{M_i}$ where
\begin{align}
  \ket{M_{j}}&=\frac{1}{\sqrt{2}}\left(\ket{01}+e^{i\phi_j}\ket{10}\right) \nonumber \\
    & j\in\{1,2,3\}, \phi_1=\pi,\phi_2=\pi/3,\phi_3=-\pi/3 \nonumber\\
  \ket{M_{k}}&=\frac{1}{\sqrt{2}}\left(\ket{00}+e^{i\phi_k}\ket{11}\right) \nonumber \\
    & k\in\{4,5\}, \phi_4=0,\phi_5=\pi \nonumber\\
  \ket{M_{l}}&=\frac{1}{{2}}\left(\ket{00}+e^{i\phi_l}\ket{01}+e^{-i\phi_l}\ket{10}+\ket{11}\right) \nonumber \\
    & l\in\{6,7,8\}, \phi_6=0,\phi_7=2\pi/3,\phi_8=-2\pi/3 \nonumber\\
  \ket{M_{9}}&=\ket{01} \nonumber \\
  \ket{M_{10}}&=\ket{10} \nonumber
\end{align}
and
\begin{eqnarray*}
  w_1=w_2=w_3 &=& 2/9 \\
  w_4 &=& 2/3 \\
  w_5 &=& 1 \\
  w_6=w_7=w_8 &=& 2/9 \\
  w_9=w_{10} &=& 1/2.
\end{eqnarray*}

The first three outcomes all project onto maximally entangled states, however the states they project into are not orthogonal, nor are they stabilizer measurements. This POVM, if used for standard BSM, would have a success probability below $50\%$ (because $(w_1+w_4+w_5)/4<1/2$). However, in terms of performing successful Type-II fusion on a state of the form Eq.~(\ref{eqn:AcrossB}), this POVM has a success probability of
\[
  \frac{w_1+w_2+w_3+w_4+w_5}{4}=7/12\approx58.3\%
\]
We have therefore demonstrated an operational distinction between Type-II fusion and BSM - the latter should be viewed as only one of many possible methods for achieving the former.\footnote{There are other protocols (such as teleportation) where an effective POVM such as this works and so is possibly preferable to the unboosted BSM.}

To understand the operation of the circuit above, it is helpful to look at the effective measurement on modes $3$,$4$ performed by the 3-DFT and detectors indicated by the dashed box. In the subspace wherein only one photon enters modes $3$,$4$ the 3-DFT projects onto unbiased states of the form $\fket{10}+e^{i\phi}\fket{01}$ with $\phi\in\{0,\pm2\pi/3\}$. In the subspace with two photons in modes $3$,$4$ the effective measurement is given by the seven Kraus operators $\sqrt{w_i}\bra{L_i}$ where
\begin{align}
  \ket{L_{j}} &= \frac{1}{\sqrt{2}}\left(\fket{20}+e^{i\phi_j}\fket{02}\right) \nonumber \\
    & j\in\{1,2,3\}, \phi_1=\pi,\phi_2=\pi/3,\phi_3=-\pi/3 \nonumber \\
  \ket{L_{k}} &= \frac{1}{{2}}\left(\fket{20}+e^{i\phi_k}\fket{02}\right)+\frac{e^{-i\phi_k}}{\sqrt{2}}\fket{11} \nonumber \\
    & k\in\{4,5,6\}, \phi_4=0,\phi_5=2\pi/3,\phi_6=-2\pi/3 \nonumber \\
  \ket{L_{7}} &= \fket{11} \nonumber
\end{align}
and 
\begin{eqnarray*}
  w_1=w_2=w_3=w_4=w_5=w_6&=&4/9 \\
  w_7&=&1/3.
\end{eqnarray*}

The important feature of this measurement is that the first three outcomes are unbiased projectors onto $\fket{20}$, $\fket{02}$ that do not have any weight on the $\fket{11}$ state. Achieving this is not possible without using a non-vacuum ancilla of some form. 

It is the assistance of the single photon ancilla that is ultimately responsible for boosting the efficacy of this particular Type-II fusion protocol over the $50\%$ barrier. Standard BSMs are not known to be boostable above $50\%$ using a single photon. 

It is, however, possible to boost BSM using two single photon ancillae. The best known such boosting protocol was given in~\cite{EL14} where the success probability achieved was $5/8=62.5\%$. To enable a direct comparison to this result, consider the case in Fig.~\ref{fig:boostedtype2} (right) where we use single photon ancilla to implement the 3-DFT measurement just described on modes $1$ and $2$ as well on modes $3$ and $4$. In that case we find we have boosted the probability of performing the Type-II fusion to $2/3$, which is above the boosted BSM probability.

%% file: figures/unboostedTypeI.tex
\centerline{\Qcircuit @C=1em @R=1em {
  \modenumber{1} & & \qw & \qw & \qw & \qw & \qw & \qw \\
  \modenumber{2} & & \qw & \control \qw & \detector & & & & \rule{3.5em}{0em}\mbox{output}\\
  & & & & & & & & \rule{3.5em}{0em}\mbox{qubit} \\
  \modenumber{3} & & \qw & \qw & \qw & \qw & \qw & \qw \\
  \modenumber{4} & & \qw & \ctrl{-3} \qw & \detector
  \inputgroupv{1}{2}{.8em}{1.1em}{\mbox{qubit 1}\qquad}
  \inputgroupv{4}{5}{.8em}{1.1em}{\mbox{qubit 2}\qquad}
  \gategroup{1}{8}{4}{8}{.75em}{\}}
}\vspace{2em}}

%% file: figures/boosted_type_1.tex
\centerline{\Qcircuit @C=1em @R=1em {
  \modenumber{1} & & \qw & \qw & \qw & \qw & \qw & \qw \\
  \modenumber{2} & & \qw & \control \qw & \detector \\
  & & & & & \\
  \modenumber{3} & & \qw & \qw & \qw & \qw & \qw & \qw \\
  \modenumber{4} & & \qw & \control \qw & \detector \\
  & & & & & \\
  \modenumber{5} & & \qw & \control \qw & \detector \\
  \modenumber{6} & & \qw & \ctrl{-6} \qw & \detector \\
  \modenumber{7} & & \qw & \qw & \qw & \qw & \qw & \qw \\
  \modenumber{8} & & \qw & \qw & \qw & \qw & \qw & \qw
  \inputgroupv{1}{2}{.8em}{1.1em}{\mbox{qubit 1}\qquad}
  \inputgroupv{4}{5}{.8em}{1.1em}{\mbox{qubit 2}\qquad}
  \inputgroupv{7}{10}{.8em}{2.1em}{\mbox{ancilla}\ket{\chi^+}\qquad\quad}
}\vspace{2em}}

%% file: figures/boostedtype2.tex
\mbox{\centerline{\Qcircuit @C=1em @R=1em { 
  & & & & & & & & & & \mbox{ancilla}\fket{1} & & & & \qw & \qw & \control\qw & \qw & \detector \\ 
  & & & & & & & & & & & & & \\ 
  \modenumber{1} & & \qw & \control \qw & \qw & \qw & \detector & & & & & & & & \ctrl{4} & \qw & \control\qw & \qw & \detector \\
  \modenumber{2} & & \qw & \qw & \control \qw & \qw & \detector & & & & & & & & \qw & \ctrl{2} & \ctrl{-3} \qw & \qw & \detector \\
  & & & & & \\
  \modenumber{3} & & \qw & \qw & \ctrl{-2} \qw & \control \qw & \detector & & & & & & & & \qw & \control\qw & \control\qw & \qw & \detector \\
  \modenumber{4} & & \qw & \ctrl{-4} \qw & \qw & \control \qw & \detector & & & & & & & & \control\qw & \qw & \control\qw & \qw & \detector \\
  & & & & & \\
  \modenumber{5} & & \qw & \qw & \qw & \ctrl{-3} \qw & \detector & & & & \mbox{ancilla}\fket{1} & & & & \qw & \qw & \ctrl{-3} & \qw & \detector
  \inputgroupv{3}{4}{.8em}{1.1em}{\mbox{qubit 1}\qquad}
  \inputgroupv{6}{7}{.8em}{1.1em}{\mbox{qubit 2}\qquad}
  \inputgroup{9}{9}{0em}{\mbox{ancilla} \fket{1}\qquad\qquad}
}}\hspace{-3em}}\vspace{1em}

%% file: sections/conclusion.tex
Primitives for quantum computing such as preparation of small entangled dual-rail
qubit states, as well as entangling multi-qubit measurements are inherently
probabilistic in linear optics. Using small examples,
we have laid out a series of techniques that are able to improve those success
probabilities significantly.

For the examples of Bell and GHZ state generators, variations range from using
ancillary vacuum modes in conjunction with elaborate feed-forward, boosting by means
of additional single photon states, to intertwining the state generation circuits
with multiplexing. It should be noted that for actual implementations, aspects
of several of the above variations may be combined in order to make the most
of the restrictions dictated by the physical platform, such as a loss budget,
cost (speed and loss) of feed-forward, and others. 
A comprehensive analysis is needed to establish appropriate trade-offs between system constraints such as footprint (physical size of the circuit), time needed for a specific task (thus limiting clock rates), energy consumption and deposition (for example due to pump processes), among others.
The scope of this paper has been to show protocols for generation of entangled
photonic states with the potential of reducing resource requirements without
analysing error propagation or discussing trade-offs due to any such errors.

The advantage of the schemes we present can be seen in that even the most na\"ive approach using a single variation is able to achieve
a substantial reduction in resources: Consider a BSG that needs to produce Bell
states with a probability of $90\%$. Building it from independent ``traditional''
BSGs succeeding with probability $18.75\%$~\cite{SLMT17}, requires about $11$ copies of a BSG.
Using a BSG with five \emph{bleeding} stages, this number shrinks to about $5$,
while not requiring more photon sources.
When multiple Bell states are combined in order to build resource states
(as needed for FBQC~\cite{PsiQ_FBQC}) such differences will become even more pronounced
and will manifest in a footprint reduction of the whole quantum computer of
roughly the same amount.
However, the cost of feed-forward, sources
of errors such as losses due to active and passive components, etc. need to be
carefully considered in any attempt to find the best ``operating point''. 

Similarly, we were able to show ways to ``boost'' success probabilities
of fusion circuits. In the realm of percolation-based resource state generation
and allowing for a local correction step, this means that arbitrarily
sized cluster states can be built starting from Bell states alone.
With fusions being a central ingredient to FBQC, ``boosting'' manifests in
substantially reduced requirements for resource state multiplexing and
thus in a lower overall footprint.

What all the results above have in common is a way of looking at linear optics
protocols less with a focus on qubits, but more on operations that are
``natural'' in linear optics. Thus one may venture to expect that
more techniques and formalisms tailored to the constraints of linear optics
may open up even more avenues to improve linear optics-based quantum
circuits.

How improvements of single stages like Bell state generation affect whole linear
optical system architectures is outside the scope of this paper.
We refer the reader to discussions of multiplexing schemes~\cite{PsiQ_SN}
and the framework of fusion-based quantum computing~\cite{PsiQ_FBQC} for more
details.

%% file: sections/acknowledgements.tex
The authors would like to thank Hector Bomb\'in, Jacob Bulmer, Hugo Cable, Axel Dahlberg, Chris Dawson, Andrew Doherty,
Megan Durney, Nicholas Harrigan, Isaac Kim, Daniel Litinski, Ye-hua Liu, Kiran Mathew, Ryan Mishmash,
Sam Morley-Short, Naomi Nickerson, Andrea Olivo, Sam Pallister, Fernando Pastawski, William Pol, Sam Roberts, Karthik Seetharam, Jordan Sullivan,
Andrzej P\'erez Veitia
and all our colleagues at PsiQuantum for useful discussions.
Terry Rudolph is on leave from Imperial College London, whose support he gratefully acknowledges.

%% file: sections/appendix.tex
\section{$H^{(4)}$-measurement}
\label{sec:h4_projection}
\begin{eqnarray}
  \bra{m'_{(1100)}} &=& -\bra{m'_{(0011)}} = 2^{-1/2}\bra{\phi^-} \label{eqn:povm1} \\
  \bra{m'_{(1010)}} &=& -\bra{m'_{(0101)}} = 2^{-1/2}\bra{\chi^-} \\
  \bra{m'_{(1001)}} &=& -\bra{m'_{(0110)}} = 2^{-1/2}\bra{\psi^-} \\
  \bra{m'_{(2000)}} &=& \sqrt{3}/2 \bra{W_{4,2}} \\
  \bra{m'_{(0200)}} &=& \sqrt{3}/2 \bra{W_{4,2}} \hat \Phi_1\hat \Phi_3 \\
  \bra{m'_{(0020)}} &=& \sqrt{3}/2 \bra{W_{4,2}} \hat \Phi_3\hat \Phi_4 \\
  \bra{m'_{(0002)}} &=& \sqrt{3}/2 \bra{W_{4,2}} \hat \Phi_1\hat \Phi_4 \label{eqn:povm2}
\end{eqnarray}
Here,
\begin{eqnarray*}
  \ket{W_{4,2}}  &=& 6^{-1/2}(\fket{1100}+\fket{1010}+\fket{1001} \\
                  && + \fket{0110}+\fket{0101}+\fket{0011})
\end{eqnarray*}
and $\hat \Phi_i$ a $\pi$-phase on detector mode $i$,
\[
  \hat \Phi_i = (-1)^{\hat n_i} .
\]
The $\ket{m'}$ are a basis for the space of two single photons distributed across four modes:
\[
  \sum_{\sum_in_i=2}\ketbra{m'_{({\bf n})}}{m'_{({\bf n})}} = \hat \id_{4,2}
\]

\section{Bleeding probabilities}
\label{sec:bleeding_probs}
For subtracting two photons from a signal of four photons in four modes (as in the Bell state generator),
there are two types of successful outcomes.
By introducing the {\it effective} reflectivity and transmissivity of stage $k$,
\begin{eqnarray*}
  r_{\rm eff}(k) &=&
    \underbrace{r_k}_{\substack{\text{reflection in}\\\text{stage $k$}}}
    \underbrace{\prod_{i=1}^{k-1}(1-r_i)}_{\substack{\text{transmission in the}\\\text{first $k-1$ stages}}}
    \qquad \mbox{and} \\
  t_{\rm eff}(k) &=& \underbrace{\prod_{i=1}^{k}(1-r_i)}_{\substack{\text{transmission in all}\\\text{$k$ stages}}} \,,
\end{eqnarray*}
we can express those outcome probabilities easily:
\begin{itemize}
  \item Detection of two photons in stage $k$. While two-photon events are completely suppressed in the limit $S\rightarrow\infty$,
    they have an effect on bleeding with a finite number of stages.
    \[
      p_2(k, S) = {4 \choose 2} r_{\rm eff}^2(k)t_{\rm eff}^2(k)
    \]
  \item Detection of one photon in stage $k_1$ and one in $k_2>k_1$
    \[
      p_1(k_1, k_2, S) = {4 \choose 1}r_{\rm eff}(k_1) {3 \choose 1}t_{\rm eff}^2(k_2)r_{\rm eff}(k_2)
    \]
\end{itemize}

Summing over all possible $k$ and $(k_1,k_2)$, we get the total probability for two-photon detections,
\[
  p(S) = \sum_{1\le k\le S}p_2(k,S) + \sum_{1\le k_1<k_2\le S}p_1(k_1,k_2,S)
\]
For reflectivities causing equal spreading between the bleeding stages~(\ref{eqn:equal_spread}),
those sums can be evaluated explicitly and we obtain
\[
  p(S) = \frac{S(1 + S + S^2)}{(1 + S)^3}\,.
\]

\section{Bleeding losses}
\label{sec:bleeding_loss}
The lowest order approximation for switch losses $r\,\epsilon/4$ (from Sec.~\ref{sec:classical_control}) is independent of the switch state,
so the same whether the switch is turned on or off. To lowest order in $\epsilon$, the total loss incurred in a bleeding circuit
would be the sum of losses in all stages, so
\[
  L(S) = \frac{\epsilon}{4} \sum_{k=1}^{S} r_k \,.
\]
In order to compare the effective loss due to switching in a bleeding scheme with the loss due to a single active phase shifter injection loss,
we are looking for the number of stages $S$ below which the single active phase shifter is more lossy,
\[
  S_1 = \max\limits_{L(S)\le\epsilon} S \,.
\]
In the case of equal spreading (Eq.~\ref{eqn:equal_spread}), $L(S)$ increases logarithmically with the number of stages and $S_1=81$.

\section{Spatial bleeding}
\label{sec:spatial_bleeding}
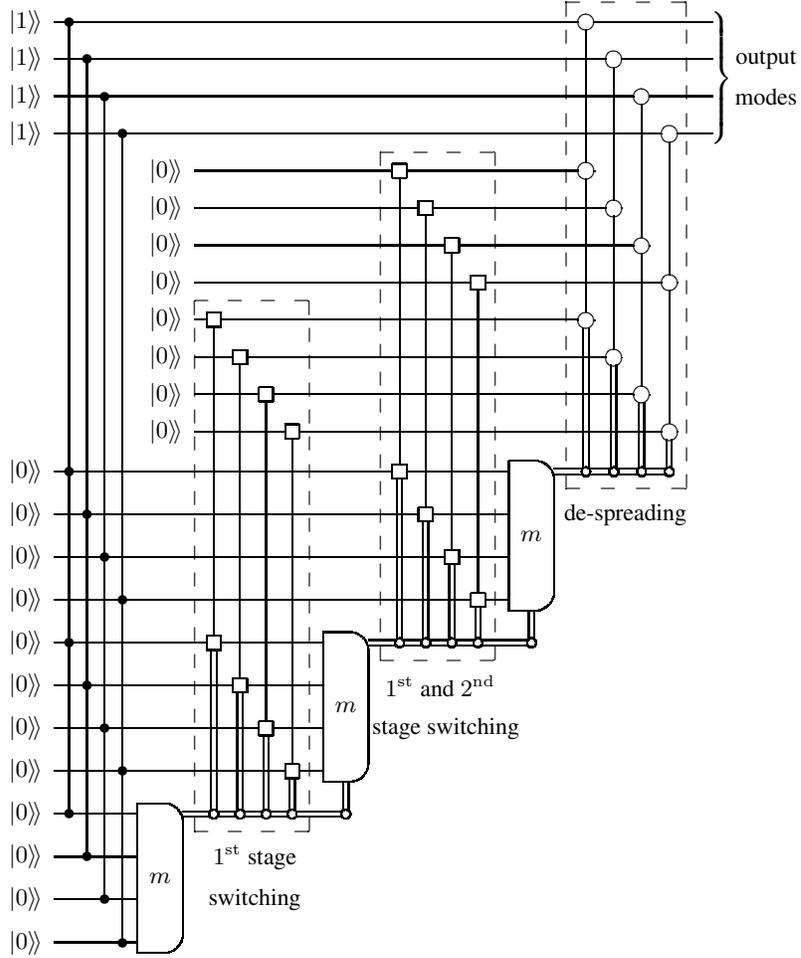
\begin{figure*}[t]
  \input{figures/spatial_bleeding.tex}
  \caption{
    A ``spatial'' implementation of the bleeding protocol as shown in Figs.~\ref{fig:bleeding_protocol} and ~\ref{fig:spatial_bleeding}
    for $S=3$ stages.
    The first four passives ``spread'' the input photons into superpositions on $4$ modes each. For an equal-amplitude spread over all stages
    (see Fig.~\ref{fig:bleeding_stages}), unbiased $S$-mode multiports (such as Hadamards or DFTs) can be used. \\
    After each stage's detections using $m$ (as defined in Fig.~\ref{fig:single_mode}), the remaining modes will be switched
    out to be de-spread if a total of two photons was detected
    (the two-mode elements with squares are switches, for example MZI-based as in Fig.~\ref{fig:mzi_subtraction} with $r=1$).\\
    The final step of ``de-spreading'', which may be an active switch-like network or a passive interferometer for probabilisitc operation,
    concentrates the photon-subtracted states back in the four signal modes. De-spreading here can be thought of as applying a 3-mode DFT to each of the four
    mode triplets if the first stage succeeded, a 2-mode DFT (50:50 beam splitter) if the second stage succeeded, or the identity if the third stage succeeded.
    \label{fig:spatial_bleeding}
  }
\end{figure*}

In Fig.~\ref{fig:spatial_bleeding} we present an example how the bleeding protocol introduced in Sec.~\ref{sec:bleeding} can be implemented by ``bleeding'' into spatially separated
modes rather than successive attempts of photon subtraction on the same signal modes.

\section{Primate Expressions}
\label{sec:primate_expression}
In this appendix we present mathematical expressions for (1) the probability of fusing two primates, (2) the post fusion primate state and (3) the probability of turning a primate into a GHZ state. We then use these expressions to compute the number of single photons required to produce a 4-GHZ state as presented in  Table~\ref{tab:four_ghz_resources}.

By backpropagating Fock states from the detectors and an identity on the output modes, we can obtain an explicit form for the fusion operation $F(t)$ as a set of Kraus operators $F(t) = \{\hat{M}_{(qr)}\}$. The elements corresponding to successful fusion are:
\[
    \hat{M}^{ij}_{(10)}, \hat{M}^{ij}_{(01)} = \frac{1}{\sqrt{2}}({\hat{a}}_i \pm {\hat{a}}_j) \left[t^{\hat{n}_{ij}-1}(1-t)\right]^{1/2}.
\]
where $\hat{n}_{ij}$ is the total photon number operator for modes $i$ and $j$.
Now we can compute the unnormalized post-fusion state to be:
\begin{widetext}
\[
  \begin{split}
  \hat{M}^{2n_1, 2n_1+1}_{(10)} \ket{\pi_{n_1}^{\lambda_1, \pm}}\!\ket{\pi_{n_2}^{\lambda_2, \pm}} =
  \sqrt{\frac{t(1-t)}{2}} \Bigg[ &
    \sqrt{\frac{\lambda_2 \lambda_2}{2}} \left(
      \fket{2}\!\fket{01}^{\otimes n_1+n_2-1}\!\fket{0} \pm \fket{0}\!\fket{10}^{\otimes n_1+n_2-1}\!\fket{2}
    \right) \\
    & + \fket{0} \left(\sqrt{\lambda_1(1-\lambda_2)} \ket{\phi_1} + \sqrt{\lambda_2(1-\lambda_1)} \ket{\phi_2} + t \sqrt{\lambda_1 \lambda_2} \ket{\phi_3}
    \right)\!
    \fket{0}
  \Bigg]
\end{split}
\]
\end{widetext}
where the Kraus operator has acted on the last mode of $\ket{\pi_{n_1}^{\lambda_1, \pm}}$ and the first mode of $\ket{\pi_{n_2}^{\lambda_2, \pm}}$ and $\ket{\phi_1}, \ket{\phi_2}, \ket{\phi_3}$ are normalized orthogonal states. Applying $\hat{M}^{ij}_{(01)}$ yields a state of the same form so we can compute the probability of obtaining either success outcome to be:
\[
  p_{\rm p}(\lambda_1,\lambda_2) = t(1-t)\left[ \lambda_1 + \lambda_2 - (1 - t^2) \right]
\]
and the normalized state is a primate $\ket{\pi_{n^\prime}^{\lambda^\prime, \pm}}$ with 
\[
  \lambda^\prime(\lambda_1,\lambda_2) = \lambda_1\lambda_2\big/\left[\lambda_1 + \lambda_2 - \lambda_1\lambda_2(1-t^2)\right]
\]
and $n^\prime = n_1 + n_2$. Similarly, we can compute the probability of turning two primates into a GHZ state by applying fusions between pairs of outermost modes, as depicted in Fig.~\ref{fig:fusing_primates}, to be: 
\[
  p_{\rm G}(\lambda_1, \lambda_2) = \lambda_1\lambda_2/8
\]
when using two fusion operations both with $t=1/2$, which maximizes the success probability for all primates. 

We can then compute the number of single photons required to produce a 4-GHZ state using the perfect MUX construction introduced in Sec.~\ref{sec:primates_multiplexing}. The number of single photons used to produce a 2-primate from deterministic 1-primates ($\lambda=1$) is:
\[
  N_2=\frac{4}{p_{\rm p}(1,1)} = \frac{4}{(t-t^2)(1+t^2)}
\]
such that the number of photons required to produce a 4-GHZ state from two 2-primates is:
$$
  N = \frac{2N_2}{p_{\rm G}(\lambda^\prime(1,1), \lambda^\prime(1,1))} = 64\frac{1+t^2}{t-t^2}
$$
where $t$ is the transmissivity used to form both of the 2-primates and the final fusions have been set to $t=1/2$. $N(t)$ is minimized with $t=\sqrt{2}-1$ yielding $N(\sqrt{2}-1) = 128(1+\sqrt{2}) \approx 309$ which is the number given in Table~\ref{tab:four_ghz_resources}.

%% file: figures/spatial_bleeding.tex
\newcommand{\cgate}[1]{*+<.6em>{#1} \POS ="i","i"+UR;"i"+UL **\dir{-};"i"+DL **\dir{-};"i"+DR **\dir{-};"i"+UR **\dir{-},"i" \cw}
\centerline{\Qcircuit @C=.5em @R=.9em {
  \lstick{\fket{1}} & \ctrl{20} & \qw & \qw & \qw   & \qw              & \qw               & \qw & \qw & \qw & \qw & \qw & \qw & \qw & \qw & \qw & \qw & \qw & \qw & \qw & \qw & \measure{}\qw & \qw & \qw & \qw & \qw & \qw & \qw \\
  \lstick{\fket{1}} & \qw & \ctrl{20} & \qw & \qw   & \qw              & \qw               & \qw & \qw & \qw & \qw & \qw & \qw & \qw & \qw & \qw & \qw & \qw & \qw & \qw & \qw & \qw & \measure{}\qw & \qw & \qw & \qw & \qw & \qw & \rule{3.5em}{0em}\mbox{output} \\
  \lstick{\fket{1}} & \qw & \qw & \ctrl{20} & \qw   & \qw              & \qw               & \qw & \qw & \qw & \qw & \qw & \qw & \qw & \qw & \qw & \qw & \qw & \qw & \qw & \qw & \qw & \qw & \measure{}\qw & \qw & \qw & \qw & \qw & \rule{3.5em}{0em}\mbox{modes} \\
  \lstick{\fket{1}} & \qw & \qw & \qw & \ctrl{20}   & \qw              & \qw               & \qw & \qw & \qw & \qw & \qw & \qw & \qw & \qw & \qw & \qw & \qw & \qw & \qw & \qw & \qw & \qw & \qw & \measure{}\qw & \qw & \qw & \qw {\gategroup{1}{28}{4}{28}{.75em}{\}}} \\
                    &     &     &     &             &                  & \lstick{\fket{0}} & \qw & \qw & \qw & \qw & \qw & \qw & \qw & \gate{}\qwx[8] & \qw & \qw & \qw & \qw & \qw & \qw & \measure{}\qwx[4]\qwx[-4] \\
                    &     &     &     &             &                  & \lstick{\fket{0}} & \qw & \qw & \qw & \qw & \qw & \qw & \qw & \qw & \gate{}\qwx[8] & \qw & \qw & \qw & \qw & \qw & \qw & \measure{}\qwx[4]\qwx[-4] \\
                    &     &     &     &             &                  & \lstick{\fket{0}} & \qw & \qw & \qw & \qw & \qw & \qw & \qw & \qw & \qw & \gate{}\qwx[8] & \qw & \qw & \qw & \qw & \qw & \qw & \measure{}\qwx[4]\qwx[-4] \\
                    &     &     &     &             &                  & \lstick{\fket{0}} & \qw & \qw & \qw & \qw & \qw & \qw & \qw & \qw & \qw & \qw & \gate{}\qwx[8] & \qw & \qw & \qw & \qw & \qw & \qw & \measure{}\qwx[4]\qwx[-4] \\
                    &     &     &     &             &                  & \lstick{\fket{0}} & \gate{}\qwx[8] & \qw & \qw & \qw & \qw & \qw & \qw & \qw & \qw & \qw & \qw & \qw & \qw & \qw & \measure{} \\
                    &     &     &     &             &                  & \lstick{\fket{0}} & \qw & \gate{}\qwx[8] & \qw & \qw & \qw & \qw & \qw & \qw & \qw & \qw & \qw & \qw & \qw & \qw & \qw & \measure{} \\
                    &     &     &     &             &                  & \lstick{\fket{0}} & \qw & \qw & \gate{}\qwx[8] & \qw & \qw & \qw & \qw & \qw & \qw & \qw & \qw & \qw & \qw & \qw & \qw & \qw & \measure{} \\
                    &     &     &     &             &                  & \lstick{\fket{0}} & \qw & \qw & \qw & \gate{}\qwx[8] & \qw & \qw & \qw & \qw & \qw & \qw & \qw & \qw & \qw & \qw & \qw & \qw & \qw & \measure{} \\
  \lstick{\fket{0}} & \control\qw & \qw & \qw & \qw & \qw              & \qw               & \qw & \qw & \qw & \qw & \qw & \qw & \qw & \gate{} & \qw & \qw & \qw & \qw & \multimeasureD{3}{m} & \cw & \controlo\cw\cwx[-4] & \controlo\cw\cwx[-3] & \controlo\cw\cwx[-2] & \controlo\cw\cwx[-1] \\
  \lstick{\fket{0}} & \qw & \control\qw & \qw & \qw & \qw              & \qw               & \qw & \qw & \qw & \qw & \qw & \qw & \qw & \qw & \gate{} & \qw & \qw & \qw & \ghost{m} & & & \rule{1em}{0em}\mbox{de-spreading}\\
  \lstick{\fket{0}} & \qw & \qw & \control\qw & \qw & \qw              & \qw               & \qw & \qw & \qw & \qw & \qw & \qw & \qw & \qw & \qw & \gate{} & \qw & \qw & \ghost{m} \\
  \lstick{\fket{0}} & \qw & \qw & \qw & \control\qw & \qw              & \qw               & \qw & \qw & \qw & \qw & \qw & \qw & \qw & \qw & \qw & \qw & \gate{} & \qw & \ghost{m} \\
  \lstick{\fket{0}} & \control\qw & \qw & \qw & \qw & \qw              & \qw               & \gate{} & \qw & \qw & \qw & \qw & \multimeasureD{3}{m} & \cw & \controlo\cw\cwx[-4] & \controlo\cw\cwx[-3] & \controlo\cw\cwx[-2] & \controlo\cw\cwx[-1] & \cw & \controlo\cw\cwx[-1] \\
  \lstick{\fket{0}} & \qw & \control\qw & \qw & \qw & \qw              & \qw               & \qw & \gate{} & \qw & \qw & \qw & \ghost{m} & & & \mbox{\rule{1em}{0em} $1^{\rm st}$ and $2^{\rm nd}$}\\
  \lstick{\fket{0}} & \qw & \qw & \control\qw & \qw & \qw              & \qw               & \qw & \qw & \gate{} & \qw & \qw & \ghost{m} & & & \mbox{\rule{1.7em}{0em}stage switching}\\
  \lstick{\fket{0}} & \qw & \qw & \qw & \control\qw & \qw              & \qw               & \qw & \qw & \qw & \gate{} & \qw & \ghost{m} \\
  \lstick{\fket{0}} & \control\qw & \qw & \qw & \qw & \multimeasureD{3}{m} & \cw & \controlo\cw\cwx[-4] & \controlo\cw\cwx[-3] & \controlo\cw\cwx[-2] & \controlo\cw\cwx[-1] & \cw & \controlo\cw\cwx[-1] \\
  \lstick{\fket{0}} & \qw & \control\qw & \qw & \qw & \ghost{m}        & & & \mbox{\rule{1em}{0em} $1^{\rm st}$ stage} \\
  \lstick{\fket{0}} & \qw & \qw & \control\qw & \qw & \ghost{m}        & & & \mbox{\rule{1em}{0em} switching} \\
  \lstick{\fket{0}} & \qw & \qw & \qw & \control\qw & \ghost{m}        \gategroup{1}{22}{13}{25}{1em}{--} \gategroup{5}{15}{17}{18}{1em}{--} \gategroup{9}{8}{21}{11}{1em}{--}
}\vspace{1em}}